\documentclass[fleqn,usenatbib]{mnras}

\usepackage{newtxtext,newtxmath, diagbox, mathtools}

\usepackage[T1]{fontenc}

\DeclareRobustCommand{\VAN}[3]{#2}
\let\VANthebibliography\thebibliography
\def\thebibliography{\DeclareRobustCommand{\VAN}[3]{##3}\VANthebibliography}

\usepackage{graphicx}	
\usepackage{amsmath}	
\usepackage{mathrsfs}  
\usepackage{gensymb}
\usepackage{ amsfonts, array, latexsym, url} 
\usepackage{dsfont}

\newcommand{\Bb}{\mathcal{B}}
\newcommand{\Dd}{\mathcal{D}}
\newcommand{\Hh}{\mathcal{H}}

\newcommand{\Ss}{\mathcal{S}}
\newcommand{\Pp}{\mathscr{P}}
\newcommand{\Ff}{\mathscr{F}}

\newcommand{\argmin}{\operatornamewithlimits{argmin}}
\newcommand{\bea}{\begin{eqnarray}}
\newcommand{\eea}{\end{eqnarray}}
\newcommand{\rect}{{\mathrm{rect}}}

\title[Towards super-resolution via iMEC]{Towards Super-resolution via Iterative multi-exposure Coaddition}

\author[Wang, Li \& Kang]{
Lei Wang,$^{1,3}$\thanks{ E-mail: leiwang@pmo.ac.cn}
Guoliang Li$^{1,3}$
and Xi Kang$^{1,2}$
\\
$^{1}$Purple Mountain Observatory, Chinese Academy of Sciences (CAS), No. 10 Yuan Hua Road, Nanjing 210034, China\\
$^{2}$Zhejiang University-Purple Mountain Observatory Joint Research Center for Astronomy, Zhejiang University, Hangzhou 310027, China\\
$^{3}$National Basic Science Data Center (NBSDC), Building No.2, 4, Zhongguancun South 4th Street, Haidian District, Beijing 100190, China
}

\date{Accepted 2022 September 13. Received 2022 September 13; in original form 2021 December 14 }

\pubyear{2022}

\begin{document}
\label{firstpage}
\pagerange{\pageref{firstpage}--\pageref{lastpage}}
\maketitle

\begin{abstract}
In this article, we provide an alternative up-sampling and PSF deconvolution method for the iterative multi-exposure coaddition. Different from the previous works, the new method has a ratio-correction term, which allows the iterations to converge more rapidly to an accurate representation of the underlying image than those with difference-correction terms. By employing this method, one can coadd the under-sampled multi-exposures to a super-resolution and obtain a higher peak signal-to-noise ratio. A set of simulations show that we can take many advantages of the new method, e.g. in the signal-to-noise ratio, the average deviation of all source fluxes, super-resolution, and source distortion ratio, etc., which are friendly to astronomical photometry and morphology, and benefits faint source detection and shear measurement of weak gravitational lensing. It provides an improvement in fidelity over the previous works tested in this paper.
\end{abstract}




%
\section{Introduction}\label{sec_Intro}
Nowadays, with the increase of digital imaging equipment and the number of exposures, amounts of observation data flood into our computers. As we narrowly focus on astronomical observation, the objects of study e.g. the stars and galaxies do not change with time. Astronomers can obtain the deep field image in one single exposure by increasing the exposure time. However, the single exposure time is usually limited by a lot of factors, e.g. the telescope scheduling, the quality of attitude control of the instrument, the excessive saturation effect of the CCD, etc. In addition, the single exposure does not help anti-alias the under-sampled extended sources with different morphology in the observation. Owing to the reasons above, people employ short multiple exposures to replace the long single exposure in some situations. The multiple-exposure technology is a common way to avoid the excessive saturation effect of the physical pixels when bright sources appear in the field of view (FOV) and reduce the requirements for the long stable operation of the recording devices. So how to restore the degraded frames to an acceptable super-resolution becomes a fundamental issue in the multiple exposures processing\citep[][]{SP02,FH2002,Fruchter2011,Yue2016}. 

On one hand, images in an optical setup are inherently blurred due to the limit of diffraction of light. The diffraction pattern of a point source is the so-called Point Spread Function (PSF). It is well-known that the observational image is degraded by the convolved PSF. Although the resolution of imaging equipment is limited by the diffraction of light, PSF deconvolution technology can improve the resolution to some extent. In fact, the purpose of PSF deconvolution is to compensate numerically for that degradation. In addition to the resolution improvement, indirect benefits of PSF deconvolution are contrast enhancement and noise reduction\citep[][]{Sage2017}. There are many methods for the PSF deconvolution as illustrated in a review\citep[][]{SP02}(hereafter {\it SP02}): the {\it Fourier-quotient} method, the {\it CLEAN} method\citep[][]{Hogbom1974}, the Bayesian approach [including {\it Landweber} method\citep[][]{Landweber1951}, {\it Richardson-Lucy} algorithm\citep{Richardson72,Lucy74,Shepp82} and {\it maximum entropy method, MEM}], the {\it wavelet-based deconvolution}, some ways to the super-resolution\citep[][]{Gerchberg1974, Hunt1994, Lauer1999,Yue2016,Symons2021} and so on. The PSF deconvolution is applicable to even the simplest optical setup, reducing financial costs and streamlining the acquisition pipeline.

 On the other hand, the spatial information of an image is recorded by the pixel arrays on the detector (CCD or CMOS). The sampling process is also known as the pixelation procedure in the imaging field, which means the flux from the PSF convolved underlying sky image is integrated into rows of discrete squares, i.e. pixels. We know those detector pixels are "bins" for capturing flux so that there is inherent area integration, namely pixelation. Due to the economical or technical limitations, many exposures in astronomy are commonly under-sampled by the undesirable detectors\citep{FH2002, Fruchter2011}. There is always a tradeoff between the spatial, spectral, and temporal resolutions for the telescope projects.
 
 In practice, an under-sampled detector inevitably blurs the high-frequency band over its detector scale. Therefore, the spatial details of the object are hidden behind the sampling interval, which is the so-called aliasing effect. The signal aliasing means that it is under-sampled. The way to restore the lost signal caused by signal aliasing is the so-called anti-aliasing or up-sampling. Up-sampling a single under-sampled exposure can not provide additional information about spatial resolution. The multiple-exposure technology (i.e. multiple sampling) is usually adopted to increase the sampling. If multiple exposures of the same scene with sub-pixel misalignment can be acquired, the complementary information between them can be utilized to reconstruct a higher-resolution image. The multi-exposures coaddition (MEC) technique is widely adopted in astronomical observation, medical imaging, and even photography, which can increase the signal-to-noise ratio (SNR) and the ability of faint source detection. Many coaddition methods have been developed to restore more details from multiple exposures, such as {\it shift-and-add}\citep[][]{Bates1980, Farsiu2004a}, {\it Drizzle}\citep[][]{FH2002}, {\it Super-Drizzle}\citep[][]{Takeda2006}, \texttt{IMCOM}\citep[][]{Rowe2011}, {\it iDrizzle}\citep[][]{Fruchter2011}, {\it SPRITE}\citep[][]{Mboula2015}, {\it fiDrizzle}\citep[][]{WL2017}, and the iterative back-projection ({\it IBP})\citep[][]{Irani1991,Symons2021}. {\it Drizzle} has become a de facto standard for the combination of images taken by the Hubble Space Telescope (HST). And some {\it Drizzle}-based methods are widely used to restore fine details of under-sampled exposures to fuse the images in the same sky field.
 
Theoretically, under-sampled multiple exposures record different parts of a signal so that the signal aliasing may be partially or totally restored (which depends on the number of multiple exposures and the critical sampling rate, i.e. Nyquist sampling, on the band-limited image) by appropriately combining them, e.g. stacking and comparing. Informally, the stacking cumulates the flux over all the exposure time, which contributes the low frequency power on the signal, e.g. the {\it shift-and-add}, {\it interlacing} or {\it Drizzle} term\citep[][]{FH2002}. While the comparison between recoveries and exposures shows the high frequency details, e.g. the correction terms in the {\it iDrizzle}\citep[][]{Fruchter2011} or {\it fiDrizzle}\citep[][]{WL2017}. 

By combining the PSF deconvolution and signal anti-aliasing, we can coadd the multi-exposures to achieve a super-resolution. Super-resolution techniques have been comprehensively summarized in several studies\citep[][]{Elad1999,Capel2003,Park2003,Farsiu2004b,Ouwerkerk2006,Tian2011,Nasrollahi2014} and reviewed by \citet[][]{Yue2016} from the perspective of techniques and applications. Among them, the regularized methods are the most popular due to their effectiveness and flexibility. Therefore, most of the recent representative articles about super-resolution techniques have focused on regularized frameworks\citep[][]{Takeda2007,Ng2007,Takeda2009,Yuan2010,Babacan2011,Su2012,Liu2014,Zhang2015}. Most previous works focus on either the PSF deconvolution for single exposure (e.g. {\it CLEAN, Landweber, Richardson-Lucy, MEM} algorithm, etc.) or on anti-aliasing (or deblending, up-sampling) in the MEC (e.g.  {\it shift-and-add}, \texttt{IMCOM}, {\it Super-Drizzle} and other {\it Drizzle}-based methods). Only a few works mention the under-sampled MEC with up-sampling and PSF deconvolution [e.g. {\it Starck-Pantin} method, equation 53 in {\it SP02}, \citet[][]{Yue2016}]. The {\it Starck-Pantin} algorithm is also referred to as the expectation-maximization method and is a form of maximum likelihood estimation (MLE). It has a difference-correction term and is sensitive to the ringing effect. Besides the  {\it Starck-Pantin}, in this work, under an iterative MEC (iMEC) framework we provide an alternative method for up-sampling and PSF deconvolution coaddition ($UPDC$), which has a ratio-correction term (i.e. {\it UPDC-RC}) and a higher convergence speed. 

Above all, we package the iMEC algorithms: {\it iDrizzle}, {\it fiDrizzle}, {\it fiDrizzleRC}, {\it Landweber}, {\it Richardson-Lucy}, {\it Starck-Pantin} and {\it UPDC-RC} into a software, named \texttt{iMECs}. In this work, we mainly discuss the algorithms in \texttt{iMECs}, while regarding {\it Drizzle} and \texttt{IMCOM} as classical references. The rest of the paper is organized as follows. An imaging model, theoretical derivation, and corresponding formulism are successively provided in Section \ref{sec_LSA}. Section \ref{sec_ST} shows the simulation test and comparison for several recovery methods. Finally, the discussions and conclusions are listed in the last section (Section \ref{sec_DAC}).
\section{Resampling and convolution of astronomical imaging}\label{sec_LSA}
The observation image records not only the flux from the objects of study but also a set of combined observational effects, e.g. blurring effects (PSF, seeing, vignetting, etc.), sampling effect (pixelation), noise effect (equipment noise, environment noise), and so on. Therefore, a reasonable imaging model should take those observational effects into account. 

\subsection{Image modeling}
Following \citet[][]{Joseph2021}, let the function $\Ff:\mathbb{R}^2\to\mathbb{R}$ represent the intrinsic, continuous surface brightness of the sky. Due to the limit of diffraction of light, the function $\Ff$ is convolved by the PSF of the telescope, $\Bb$ (with a kernel $b$), when the light goes through the aperture, thus $\Bb\{\Ff\}$. The telescope is equipped with a camera with $M\times N$ pixels, which then integrates the continuous field $\Bb\{\Ff\}$ over the pixel area and samples it at the pixel positions ${(X_{m}, Y_{n})}\ \forall {(m \in [\![0, M]\!], n \in [\![0, N]\!])}$. Then we have an ideal digital image $\Ss^h\{\Bb\{\Ff\}\}$, where $h$ is the linear size of a square pixel. The function $\Ss^h$ is the 2-dimensional rectangle function defined by:

\begin{eqnarray}
\forall (x,y,r) \in \mathbb{R}^3, & \nonumber \\
\Ss^h(x, y) & =  \rect^h(x)\cdot \rect^h(y), \nonumber \\
\rect^h(r) & = 
  \begin{cases}
    1       & \quad \text{if } |r| \le h/2, \nonumber\\
    0  & \quad \text{otherwise.} \nonumber
  \end{cases} \label{eq:imaging}
\end{eqnarray}
Operation by $\Ss^h$, followed by down-sampling at the pixel centers, is identical to the integration of the convolved surface brightness $\Bb\{\Ff\}$ over the pixel's area. $\Ss^{\ast h}$ is the adjoint to $\Ss^h$, i.e. up-sampling to infinitesimal pixels. Note that $\Ff \neq \Ss^{\ast h}\{\Ss^h\{\Ff\}\} \forall (h>0)$, because the down-sampling operation erases the spatial details under a pixel scale.

By considering the noise in the acquisition process, we eventually have an observation frame:
\begin{eqnarray}\label{model}
g(X_{m}, Y_{n}) &=&\Ss^h\{\Bb\{\Ff\}\}(X_{m}, Y_{n}) + \mathcal{N}(X_{m}, Y_{n}) \nonumber \\
   &=&\Hh\{\Ff\}(X_{m}, Y_{n}) + \mathcal{N}(X_{m}, Y_{n}),
\end{eqnarray}
where $\mathcal{N}(X_{m}, Y_{n})$ describes the noise contamination on each pixel of the convolved surface brightness. For convenience of expression, following \citet[][]{Yue2016}, we rewrite $g$ by substituting the combined operation of $\Ss^h\{\Bb\{\cdot \}\}$ by ${\Hh} \{\cdot\}$ on the second equal sign. 

In observation and image processing, the image is discrete finite-dimensional, not continuous. In order to test or compare recovery methods, people often use a high-definition (well-sampled) image as an original or ground truth \citep[][]{Mboula2015} and add a set of blurring effects, geometric distortion, shift, rotation, pixelation, relative motion, noise, etc., to mimic the imaging acquisition.

For the purpose of the reference, we down-sample the continuous distribution $\Ff$ to a discrete finite-dimensional sample:
\begin{equation}\label{reference_image}
f (x_{u}, y_{v})= \Ss^w \left\{ \Ff \right\}(x_{u}, y_{v}),
\end{equation}
 where $w$ is the linear size of a grid on the reference image ($w \ll h$). While ${(x_{u}, y_{v})}\ \forall {(u \in [\![0, U]\!], v \in [\![0, V]\!])}$ are the positions of the reference grids. To improve the optical and image resolution, one should choose the size of grid $w$ to meet at least the Nyquist sampling for the super-resolution. In other words, the image $f (x_{u}, y_{v})$ should be sampled from continuous to discrete without signal aliasing.
 
Define the down-sampling operator  $\Dd^h\{\cdot \}$ and the up-sampling one $\Dd^w\{\cdot \}$ respectively as 
 \begin{eqnarray}\label{down_up}
 \Dd^h\{f\}(X_{m}, Y_{n}) \equiv \Ss^h\{ \Ss^{\ast w}\{f\} \}(X_{m}, Y_{n}) \  \rm {and} \nonumber \\
  \Dd^w\{g\}(x_{u}, y_{v}) \equiv \Ss^w\{ \Ss^{\ast h}\{g\} \}(x_{u}, y_{v}).
 \end{eqnarray}
 
Therefore, to recover the $f(x_{u}, y_{v})$ from observation image $g(X_{m}, Y_{n})$, the combined effects of ${\Hh} \{\cdot\}$ in equation \ref{model} needs to be inverted. Since the noise contamination makes the image recovery an ill-posed problem, in the absence of prior knowledge about the mechanism of noise, the MLE which measures the similarity between the observation $g$ and the current estimate $\Hh\{f\}$, will be a simple and effective approach, so that one should minimize a cost function $E(f) =\frac{1}{2}\|g-\Hh\{f\}\|^2$,
\begin{equation}\label{MLE}
\hat{f} = \argmin_{f \in \mathbb{R}} \{ E(f)  \}.
\end{equation}
Naively, we may solve equation \ref{MLE} in the Fourier domain, called the naive solution to the least squares or sometimes the Fourier-quotient method, which leads to fast computations. However, the $\Hh$ in the Fourier domain should be taken very carefully to avoid divisions by zero, which commonly results in spurious high-frequency oscillations and noise amplifications. Therefore, in the presence of noise, this method has little value in use. 

\subsection{Reconstruction methods}
In practice, people are much more concerned about a stable and acceptable solution to the problem. Therefore, instead of expressing the solution through direct inversion, here we employ an iterative gradient-descent approach to find an asymptotic solution\citep[][]{SP02,Yue2016,Sage2017}:
\begin{equation}\label{GRLSL}
f^{(i+1)}  = \Pp_C\left\{f^{(i)}+\gamma \Hh^\ast \left\{ g-\Hh\{f^{(i)}\} \right\}\right\},
\end{equation}
where $f^{(i)}$ is an approximation to the original (or reference) image $f$ in the $i$-th iteration. $\Hh^\ast$ denotes the adjoint matrix of $\Hh$. $\gamma$ is called as a step size parameter, $\Pp_C\{\cdot\}$ is the component-wise projection operator that enforces our set of constraints on the recoveries (see the variables in Table \ref{tab:params}). Obviously, the target image updates itself by adding a correction term on each iteration so that the correction term is equivalent to the difference between $f^{(i+1)}$ and $f^{(i)}$, i.e. a difference-correction. The number of iterations $i$ plays an important role in the iterative methods, which is made to act as a $pseudo-regularization$ parameter. 

Besides the iterative solution with a difference-correction term in the flux domain, one may minimize the least-squares in logarithm when all the pixels in the observation image have positive flux $g(X_{m}, Y_{n})> 0$. By using the multiplicative gradient-descent approach, we write
 \begin{equation}\label{GRLSSLE}
f^{(i+1)} = \Pp_C\left\{f^{(i)} \times \Bigg{(} \Hh^\ast \left\{ \frac{g}{ \Hh\{ f^{(i)} \} } \right\} \Bigg{)}^\gamma \right\},
\end{equation}
where the $\times$ stands for a component-wise multiplication. $\gamma$ represents the number of component-wise multiplications of the object in its bracket. Comparatively, the correction term is a ratio between two adjacent iterations, e.g. $f^{(i+1)}$ and $f^{(i)}$. Namely, it is a ratio correction that updates the target in each iteration. Equations (\ref{GRLSL} and \ref{GRLSSLE}) are called as the iterative solution to the MLE. In this work, we consider three blur cases for $\Hh \{\cdot\}$: PSF convolution $\Bb \{\cdot\}$, image resampling $\Dd^h \{\cdot\}$ and the combination of them $\Dd^h \{\Bb \{\cdot\}\}$.

\subsection{PSF deconvolution}\label{sec_PD}
By substituting the combined operation of ${\Hh} \{\cdot\}$ by the PSF convolution ${\Bb} \{\cdot\}$ in equation \ref{GRLSL} and \ref{GRLSSLE}, one may find they are coincidentally the {\it Landweber} method\citep[][]{Landweber1951} and the {\it Richardson-Lucy} algorithm\citep[][]{Richardson72, Lucy74, Meinel1986}. The main advantage of the {\it Richardson-Lucy} algorithm stems from its nonlinear nature, allowing one to recover the high-frequency components of the true images which are most affected by blur\citep[][]{Donoho95, Chen01, Elad06, Donoho06, Candes06}. This method is commonly used in astronomy. Flux is preserved and the solution is always positive. It is well-known that {\it Richardson-Lucy} algorithm is good at dealing with Poisson imaging, while {\it Landweber} method suits images with Gaussian noise. According to the above derivation, the {\it Richardson-Lucy} algorithm is actually the {\it Landweber} method in logarithm space. The ratio-correction term usually allows the {\it Richardson-Lucy} algorithm to converge faster and obtain a higher peak-signal-to-noise ratio (PSNR)\footnote{PSNR$=20\times {\rm log}_{10}(\frac{\rm Peak\ Signal\ Value}{\rm Root\ Mean\ Square\ Error})$ [dB].} than that with a difference-correction term, e.g. the {\it Landweber} method (see Figure \ref{PSNRA}).

\subsection{Resampling} \label{sec_MEC}
We know that one can reduce the aliasing effect by properly resampling the multiple exposures to a high-definition grid, i.e. up-sampling. Assuming observers get $L$ coarse (or under-sampled) exposures $\{g_1, g_2... g_L\}$ for the same source, and $g_k$ ($k=1 ... L$) is obtained by the $k$-th exposure. Define $\beta_k\equiv h/w$ to be a down-sampling factor which stands for a ratio of the sampling rate of a fine grid (pixel size $w$, e.g. the target grid) to the sampling rate of a coarse grid (pixel size $h$, e.g. the under-sampled CCD grid), thus $\beta_k\ge 1$. Let $\Dd^h_k$ represent the operation of down-sampling a fine (or high definition) image ($U \times V$) to a coarse one ($M \times N$) with a shift\footnote{$\Delta$ is a position shift array including shifts along $x$ and $y$ axis, and a rotation angle between the fine and coarse grid.}, $\Delta_k$. Thus $\Dd^w_k$ up-samples the coarse image $g_k$ to the fine grid with a shift, $-\Delta_k$. Commonly, we coadd multiple exposures from the same instrument. In this case, all down-sampling factors $\beta_k$ are the same (hereafter $\beta_k$ does not vary according to the sequence number of frame $k$, thus using $\beta$ for simplicity). According to the equation \ref{GRLSL}, one gets

\begin{eqnarray}\label{WLMDC}
f^{(i+1)} &=& f^{(i)} + \frac{1}{L_{\rm E}}\sum_{k=1}^{L} \Ss^w_k\left\{ \Ss^{\ast h}_k \left\{ {g_k}-{ \Ss^h_k\{ \Ss^{\ast w}_k\{f^{(i)}\} \} } \right\} \right\} \nonumber \\
&=& f^{(i)} + \frac{1}{L_{\rm E}}\sum_{k=1}^{L} \Dd^w_k \left\{ {g_k}-{ \Dd^h_k\{ f^{(i)} \} } \right\}.
\end{eqnarray}
On the second equal sign, we use the equation \ref{down_up} to simplify the form of the expression. Hereafter, we use this simplification in the rest of this article. $\frac{1}{L_{\rm E}}\sum_{k=1}^{L} \Dd^w_k \left\{  \cdot \right\}$ stands for up-sampling and stacking operations, i.e. {\it shift-and-add}, {\it interlacing} or {\it Drizzle}\citep[][]{FH2002}, which stacks $L$ layers of coarse exposures to a fine grid and scales the flux. Considering the fractional pixel overlap, $L_{\rm E}$ represents the times (or total area\footnote{In the unit of a fine pixel area.}) of a fine grid overlapped by the stacked exposures, i.e. the effective layers of exposures, $L_{\rm E} \le L$. Note that $L_{\rm E}$ is not necessarily an integer because some fine grids may be partially overlapped by the coarse pixels located at the edge of exposures. Therefore, $L_{\rm E}$ depends on the position relative to the pixels of exposures. Ordinarily, a $Drizzled$ image is adopted as the initial condition of the iteration, i.e. $f^{(0)}=\frac{1}{L_{\rm E}}\sum_{k=1}^{L} \Dd^w_k \left\{  g_k \right\}$. 

\begin{figure*}
\centering
\includegraphics[width=7in]{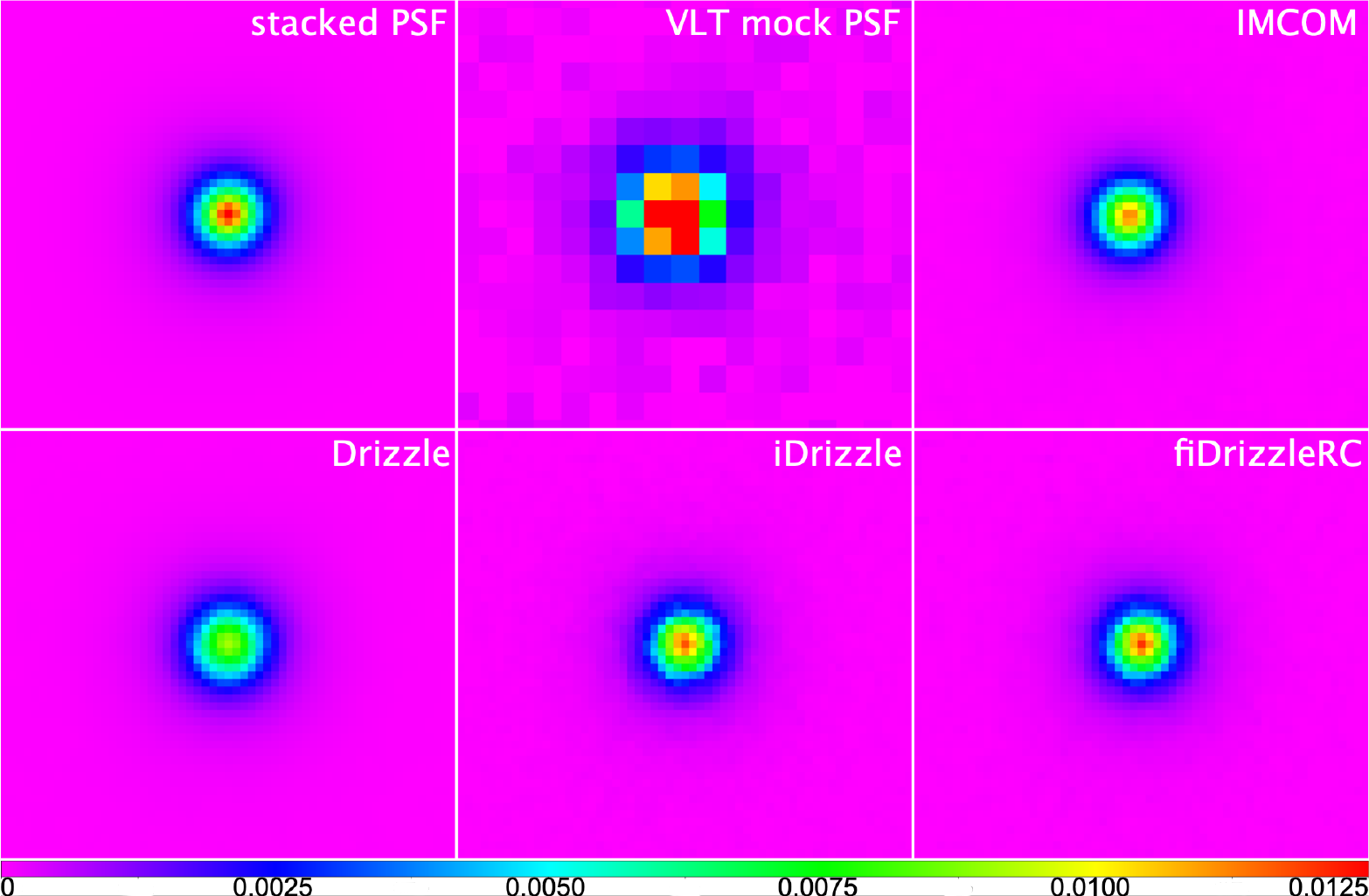} 
\caption{PSFs in different cases. Top left panel: {\it stacked} PSF, 110 PSF kernels $b_k$ are stacked pixel by pixel and normalized; top middle panel: an under-sampled PSF from one of the 110 VLT mock frames $g_k$; top right panel: a PSF coadded by $\texttt{IMCOM}$; bottom left: the {\it Drizzled} PSF; bottom middle: a PSF recovered by ${\it iDrizzle}$; bottom right: a PSF coadded by ${\it fiDrizzleRC}$.}
\label{PSFs}
\end{figure*}

\begin{figure}
\centering
\includegraphics[width=3.5in]{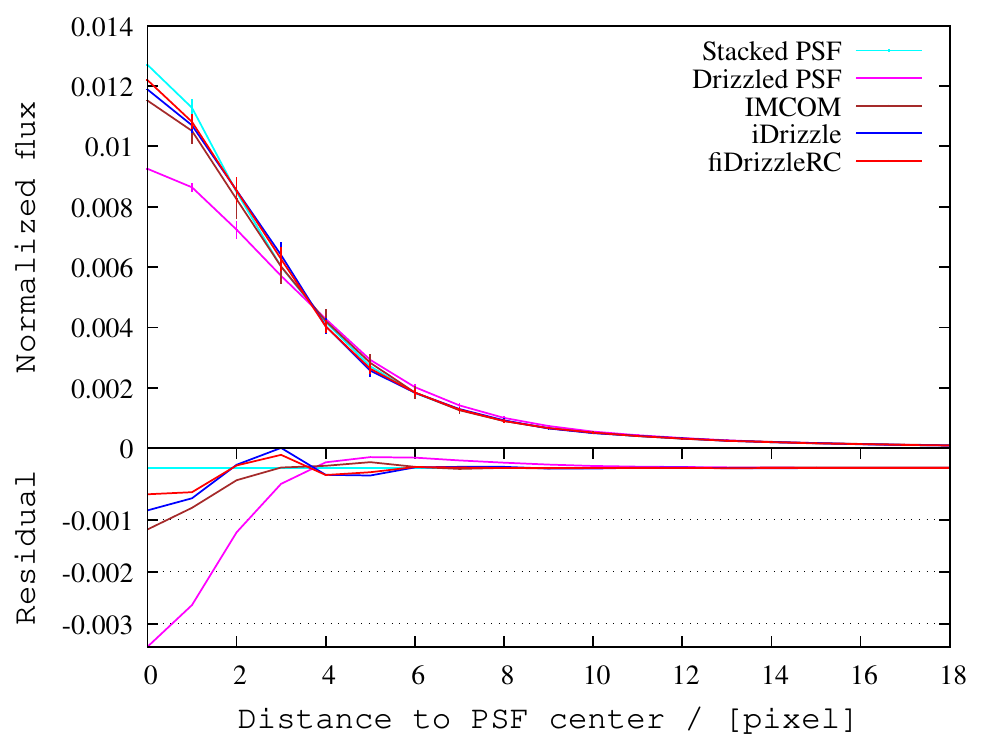} 
\caption{The PSF profiles. Top panel: the normalized PSF; bottom panel: the residuals from the {\it stacked} PSF.}
\label{PSFprofile}
\end{figure}

To up-sample a series of coarse exposures, we map the coarse pixels
 \begin{enumerate}
 \item to an over-sampled grid (beyond the Nyquist sampling) with a low-pass filter\footnote{A smooth performed in Fourier space} on the difference-correction term and finally $sinc$ interpolate to a target resolution. This algorithm is called the iterative {\it Drizzle}, namely {\it iDrizzle}\citep[][]{Fruchter2011};
\item or to a critical (or Nyquist) sampling without filter or interpolation, which is called a fast iterative {\it Drizzle}, i.e. {\it fiDrizzle}\citep[][]{WL2017}. {\it fiDrizzle} neither over-samples the exposures nor smooths the difference-correction term. Therefore, {\it fiDrizzle} avoids the effect of high sampling caused decelerating convergence \citep[][]{WL2017} and Fourier transforms (including the inverse transform) so that saves more computation than {\it iDrizzle}.
\end{enumerate}
For generalization, the difference-correction term in equation \ref{WLMDC} can be multiplied by the step size parameter $\gamma$. If all the pixels on the coarse exposures $g_k$ have positive flux, the iteration with a ratio correction also works well in the iMEC up-sampling. Under this circumstance,  a general form to coadd the multiple exposures onto a high-resolution grid is represented as
\begin{equation}\label{GWLM}
f^{(i+1)} = f^{(i)} \times \Bigg{(} \frac{1}{L_{\rm E}} \sum_{k=1}^{L} \Dd^w_k \left\{ \frac{g_k}{ \Dd^h_k\{ f^{(i)} \} } \right\} \Bigg{)}^\gamma.
\end{equation}
We call equation \ref{GWLM} as the "fast iterative {\it Drizzle} with a ratio-correction"({\it fiDrizzleRC}). Generally, {\it fiDrizzleRC} performs as well as the {\it fiDrizzle} and {\it iDrizzle} (see Figure \ref{PSFs} and \ref{PSFprofile}) in the simulation test. 

Apart from the PSF, due to the aliasing effect, an ideal point source is represented by a detector pixel on the CCD (or CMOS). For multiple exposures to the same point source, up-sampling and stacking the coarse pixels of exposures to a fine grid cell forms a bright spot (flux spreading within limited space), called the pixelation blur. When the number of stacked exposures is finite, the brightness distribution of the spot depends on the location of the point source. Different grid cell corresponds to different cases of pixel stacking. Therefore the pixelation blur is not homogeneous, even not continuously changing on the target image. Fortunately, the iMEC methods, e.g. {\it iDrizzle}, {\it fiDrizzle}, {\it fiDrizzleRC}  and the iterative {\it IBP}\citep[][]{Symons2021} are able to up-sample or anti-alias the under-sampled exposures. Deserved to be mentioned, that the difference in pixelation blur at different locations vanishes when the number of stacked exposures goes sufficiently large or to infinity. Under the circumstances, the pixelation blur has a unique form which means an ordinary PSF with a note spike (see Appendix A for detail) profile. By using the note spike profile, one may deblur the pixelation blur on the target image like deconvolving the ordinary PSF (e.g. {\it stack-and-deconvolve}). The pixelation blur is negligible only when the size of PSF is much larger than that of pixelation blur. Worth mentioning at this time is that \citet[][]{Hunt1994} achieves super-resolution on a single exposure by combining the $maximum\ a\ posteriori$ Poisson algorithm with an interpolation. So in a sense, it implies that up-sampling and PSF deconvolution for the iMEC can be handled together.

\subsection{Combination of up-sampling and PSF deconvolution}\label{sec_UPDC}
The up-sampling and PSF deconvolution are the necessary techniques to achieve the super-resolution: the up-sampling increases the definition of grids, while PSF deconvolution improves spatial resolution beyond the classical optical diffraction limit\citep[][]{Yue2016}. As mentioned before, $\Hh$ can be a combination of a series of image operations, e.g. down-sampling and PSF convolution. Let $\Hh\{ \cdot \}=\Dd^h\{ \Bb\{ \cdot \} \}$ represent convolving a target with a normalized PSF kernel $b$, after that down-sampling the PSF blurred target to a coarse grid with a down-sampling factor $\beta$. While the reverse operation $\Hh^\ast\{ \cdot \}=\Bb^\ast\{ \Dd^w\{ \cdot \} \}$ stands for up-sampling before deconvolving\footnote{In linear algebra, if $A^\ast$ and $B^\ast$ represent the adjoints of matrices $A$ and $B$ respectively, then we have $(AB)^\ast=B^\ast A^\ast$.}. According to the definition of the operator $\Dd^h\{ \Bb\{ \cdot \} \}$, this method requires the normalized PSF kernels $b_k$ have the same spatial resolution\footnote{The PSF kernel $b_k$ can generally be derived on the fine $M \times N$ grid using a set of point sources or using optical modeling of the instrument.} as the fine (target) grid on which $f$ is mapped. The {\it UPDC} for multiple exposures includes two categories according to the kinds of correction term.
\begin{enumerate}
\item One can employ the iterations with a difference correction in the iMEC. Then applying $\Hh\{ \cdot \}=\Dd^h\{ \Bb\{ \cdot \} \}$ to equation \ref{GRLSL}, we get
\begin{equation}\label{WLMPHDC}
f^{(i+1)} = f^{(i)} + \frac{1}{L_{\rm E}}\sum_{k=1}^{L} \Bb^\ast_k \left\{ \Dd^w_k \left\{ g_k- \Dd^h_k\{ \Bb_k\{f^{(i)}\}  \}  \right\} \right\},
\end{equation}
where $\frac{1}{L_{\rm E}}\sum_{k=1}^{L} \Bb^\ast_k \left\{ \Dd^w_k \left\{ \cdot  \right\} \right\}$ represents three-step operations in a sequence: up-sampling, convolving PSF and stacking, which is a method proposed by \citet[][]{SP02}. Note that, for simplicity, here $\gamma$ is set to be unity and the constraint $\Pp_C\{\cdot\}$ is neglected. We call equation \ref{WLMPHDC} as the {\it Starck-Pantin} method. In fact, if a fine (high resolution) PSF can be well measured, {\it Starck-Pantin} is able to recover the under-sampled multi-exposures to high fidelity. 

\item On the other hand, one may adopt the iterations with a ratio correction in the iMEC if flux on each pixel of $g_k$ is positive. Similarly, if the corresponding fine PSF kernels\footnote{For example, it has the same pixel scale as the high-definition image $f$.} $b_k$ are given, by applying $\Hh\{ \cdot \}=\Dd^h \{ \Bb\{ \cdot \} \}$ and $\Hh^\ast\{ \cdot \}=\Bb^\ast\{ \Dd^w\{ \cdot \} \}$ to equation \ref{GRLSSLE}, we get
\begin{equation}\label{WLMPHRC}
f^{(i+1)} =f^{(i)} \times \frac{1}{L_{\rm E}}  \sum_{k=1}^{L} \Bb^\ast_k \left\{ \Dd^w_k \left\{ \frac{g_k}{ \Dd^h_k\{ \Bb_k\{f^{(i)}\}  \} } \right\} \right\}.
\end{equation}
Here we call equation \ref{WLMPHRC} as {\it UPDC} with a ratio-correction term ({\it UPDC-RC}). The {\it Starck-Pantin} and {\it UPDC-RC} are suitable for instruments with well-modeled PSF, e.g. space telescopes. Of course, if one can measure the fine PSFs on each exposure well by using a set of stars \citep[][]{Nie2021a,Symons2021} or even galaxies \citep[][]{Nie2021b} nearby the objects, the two methods are able to provide an amazing effect on the image recovery (see the next section).
\end{enumerate}

Apparently, the methods with difference-correction terms have a wider range of application environments than those with ratio-correction terms because the subjects $g$ may have a larger domain of definition, which can make use of those images with negative flux (e.g. due to the background subtraction). However, according to the background addition and subtraction, we can also take advantage of the methods with a ratio correction, e.g. the examples in Section \ref{sec_ST}. Readers may derive the version with the step size parameter and the constraints $\Pp_C$ in equations (\ref{WLMPHDC} and \ref{WLMPHRC}). 

\begin{figure*}
\centering
\includegraphics[width=7in]{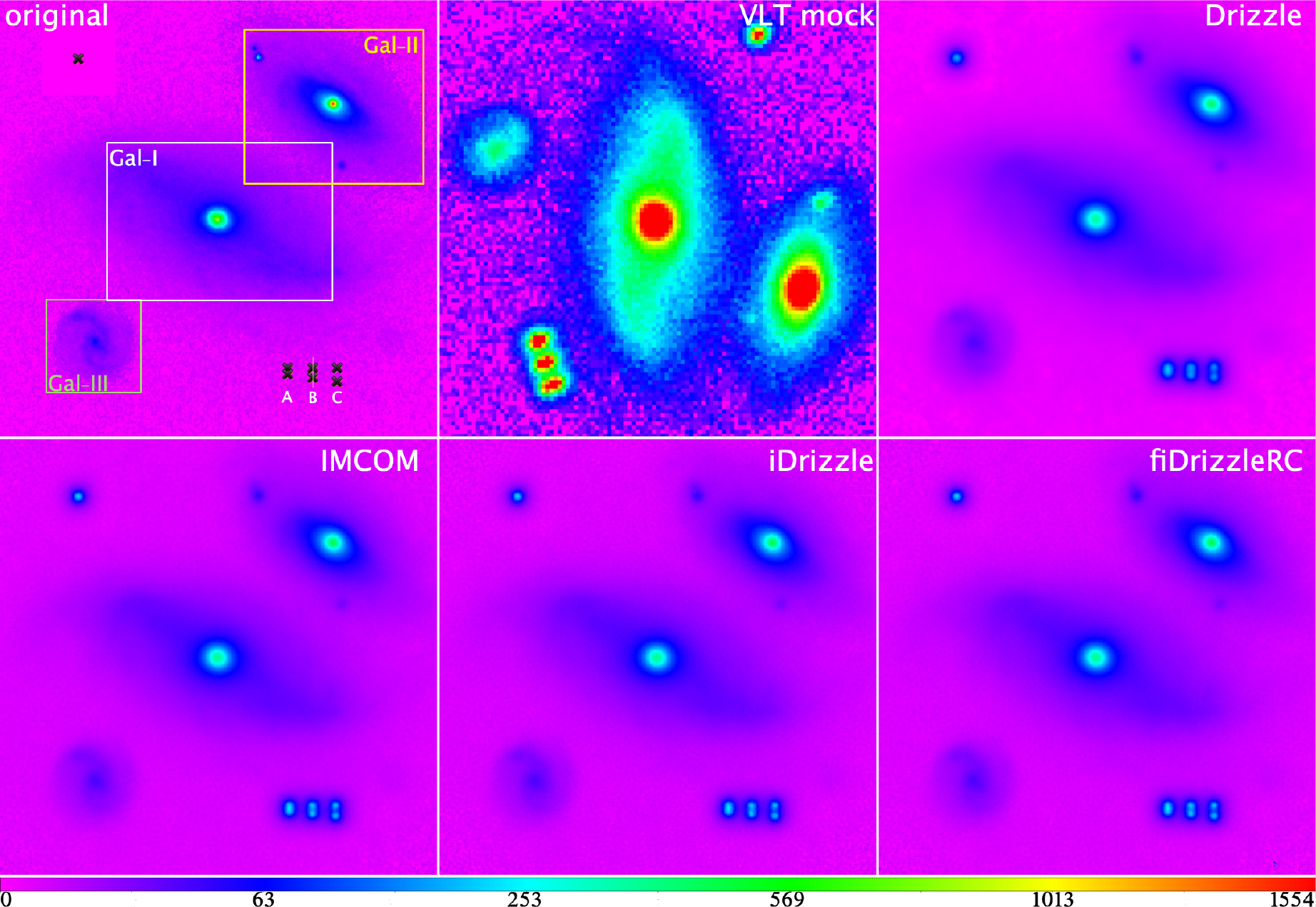} 
\caption{Image mock and up-sampling. Top left panel: the original image $f$, three binaries are shown on the bottom right, i.e. binaries {\bf A}, {\bf B} and {\bf C}. One cross stands for one pixel; top middle panel: one of the 110 VLT mock frames $g_k$, i.e. the original image $f$ sampled at VLT resolution, convolved by the VLT PSF and noise (binned by $\beta=5.25$); top right panel: the Drizzled image $f^{(0)}$; bottom left: coadded by $\texttt{IMCOM}$; bottom middle: recovered by ${\it iDrizzle}$; bottom right: coadded by ${\it fiDrizzleRC}$. Due to the coarse pixels in the middle panel, it has a larger flux around the sources. All panels have the same color bar and are scaled by the square root.}
\label{mockfig}
\end{figure*}

\begin{table}
\caption{Important Variables Used in the Text}\label{tab:params}
    \centering
    \begin{tabular}[width=\columnwidth]{p{2.1cm}||p{6cm}}
    \hline
    \bf{Variable} & \bf{Description} \\
    \hline
     $\Ff$ & intrinsic, continuous surface brightness of the sky\\
     $g_k$ & the $k$-th exposure\\
     $f$ & fully-sampled high resolution (or definition) image\\
     \hline
     $\Bb\{\cdot\}$ & PSF blurring operator, with a kernel $b$\\
    $\Ss^h\{\cdot\}$ or $\Ss^w\{\cdot\}$ & sampling operator, with pixel size $h$ or $w$\\
    $\Dd^h\{\cdot\}$ or $\Dd^w\{\cdot\}$ & down-sampling or up-sampling operator \\
    ${\Hh} \{\cdot\}$ & combined operator of blurring and sampling \\ 
    $\frac{1}{L_{\rm E}}\sum_{k=1}^{L} \Dd^w_k \left\{  \cdot \right\}$ & {\it shift-and-add}, {\it interlacing} or {\it Drizzle} operator\\
    $\Pp_C\{\cdot\}$ & component-wise projection operator for constraints set\\
    \hline
    	$\beta\equiv h/w$ &  down-sampling factor\\
	$\Delta_k$ &  shift and rotation of the $k$-th exposure\\
	$\gamma$ &  step size parameter\\
	${\langle \delta F \rangle}$ &  average deviation of all source fluxes\\
	$SDR$&source distortion ratio\\
	$PSNR$&peak-signal-to-noise ratio\\
	$SSIM$&structural similarity index measure\\
    \hline
    \end{tabular}
\end{table}

\begin{figure*}
\centering
\includegraphics[width=7in]{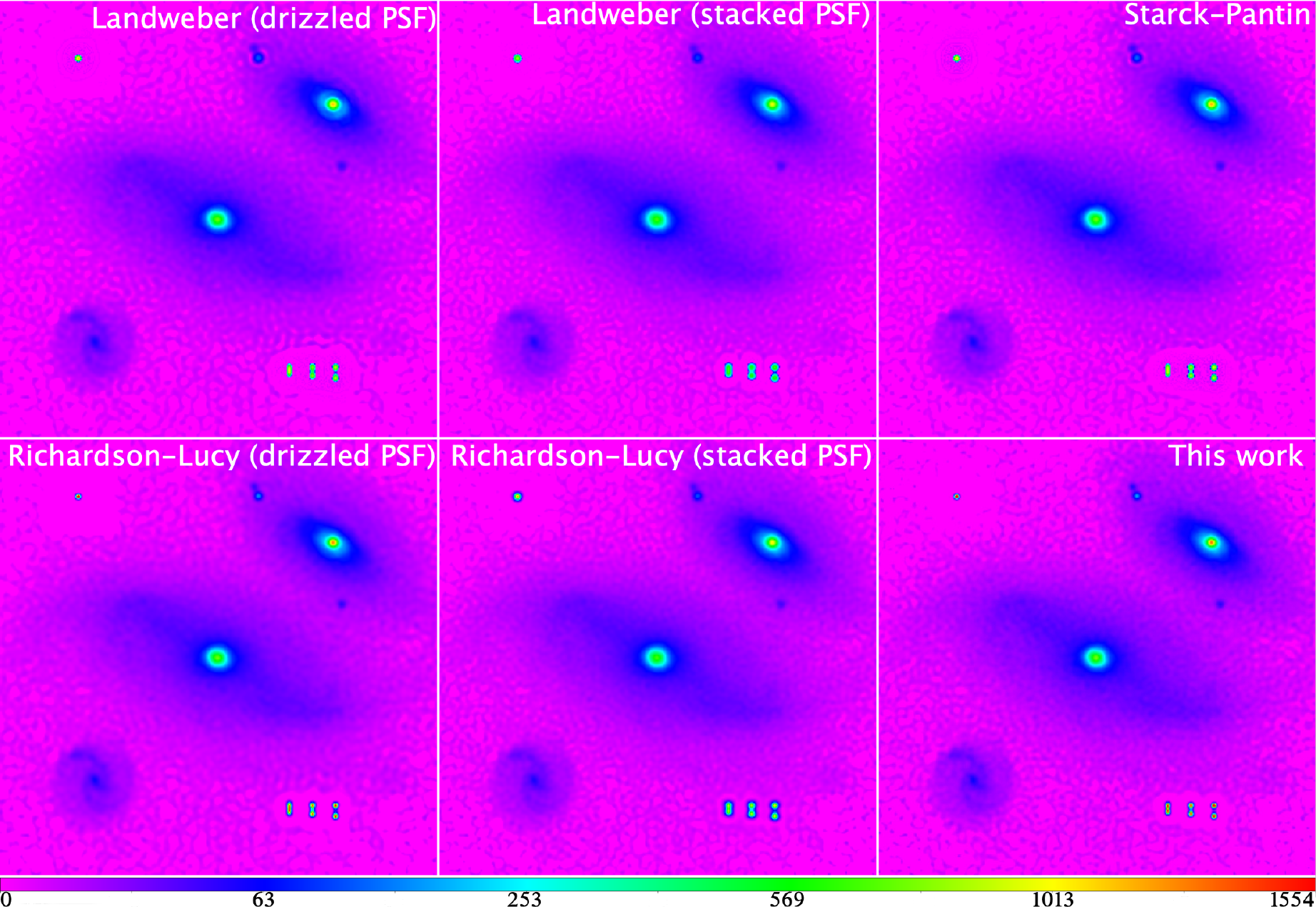} 
\caption{Recoveries to the 110 VLT mock frames. The top row images are from the iterations with a difference-correction  item (from left to right: {\it Landweber (drizzled PSF)}, {\it Landweber (stacked PSF)} and {\it Starck-Pantin}), while the bottom one from those with a ratio-correction term, from left to right: {\it Richardson-Lucy (drizzled PSF)}, {\it Richardson-Lucy (stacked PSF)} and {\it UPDC-RC}. The parameter $O_{\rm iter}$ is determined by the PSNR calculated from the pixels in the white rectangle (${\bf Gal-I}$) of Figure \ref{mockfig}.}
\label{recoveryfig}
\end{figure*}

\begin{figure*}
\centering
\includegraphics[width=7in,angle=0.0]{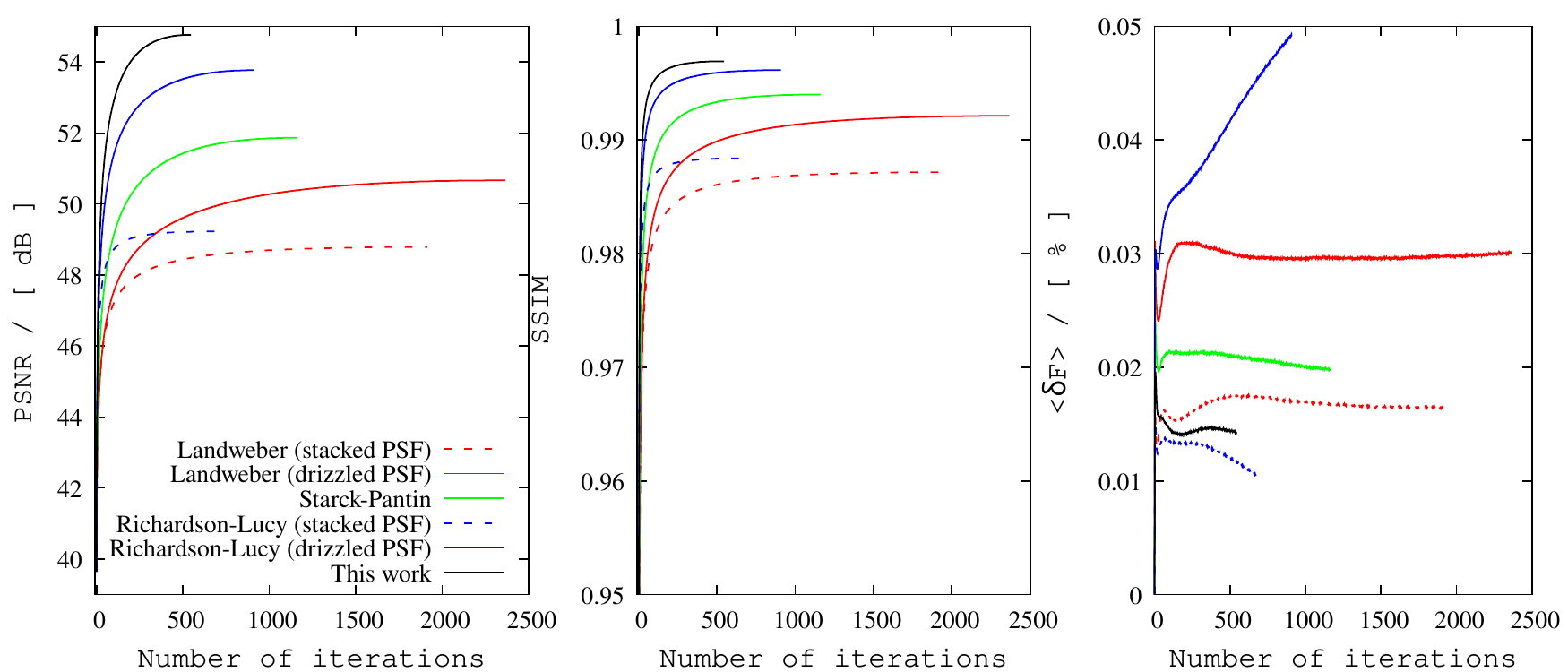} 
\caption{The diagnostic parameters (i.e. PSNR for left panel, SSIM for middle, ${\langle \delta F \rangle}$ for right) vary according to the iteration number and stop on the  $O_{\rm iter}$ which corresponds to the peak of each curve in the left panel.}
\label{PSNRA}
\end{figure*}

\begin{figure*}
\centering
\includegraphics[width=6.5in,angle=0.0]{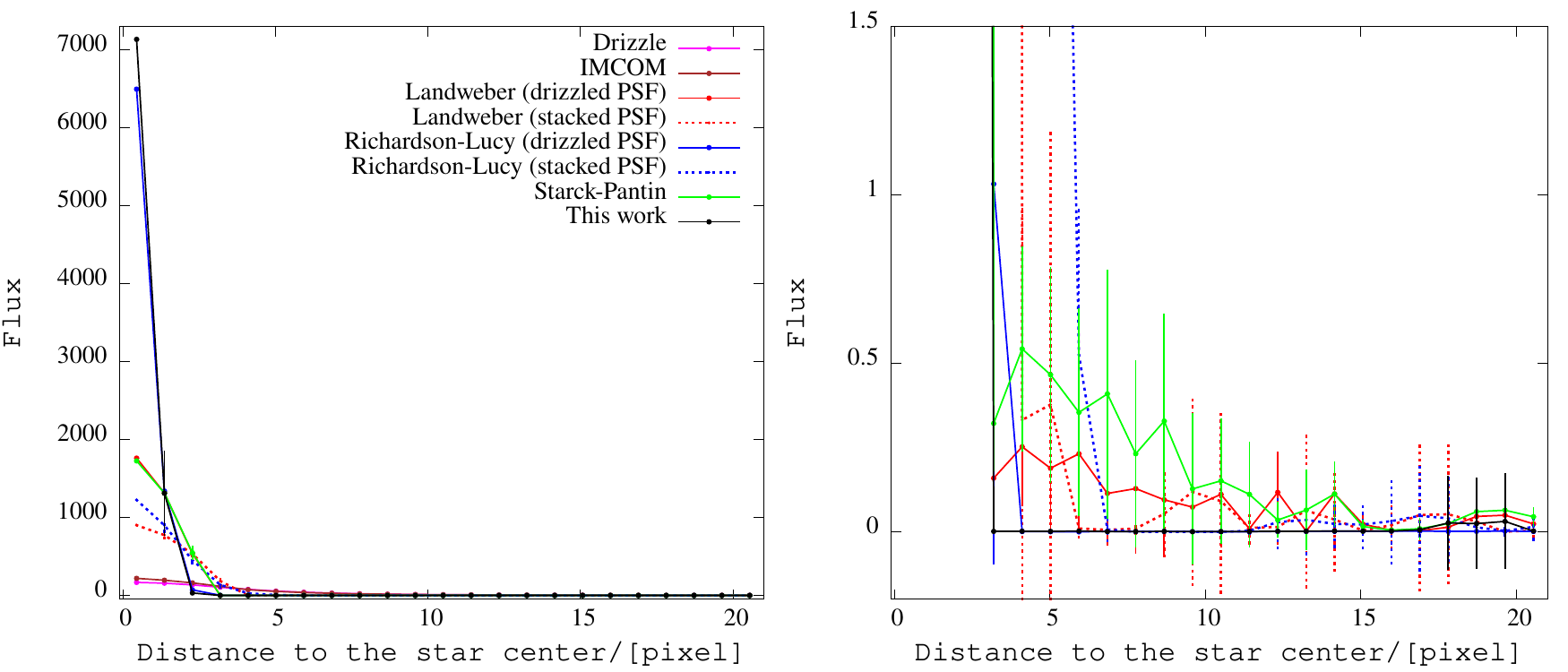} 
\caption{The star profiles. The star profiles from different recoveries are plotted on the left panel with different line types and colors. While the right panel is a zoom-in version nearby the zero flux.}
\label{starprofile}
\end{figure*}

\begin{figure}
\centering
\includegraphics[width=3.5in]{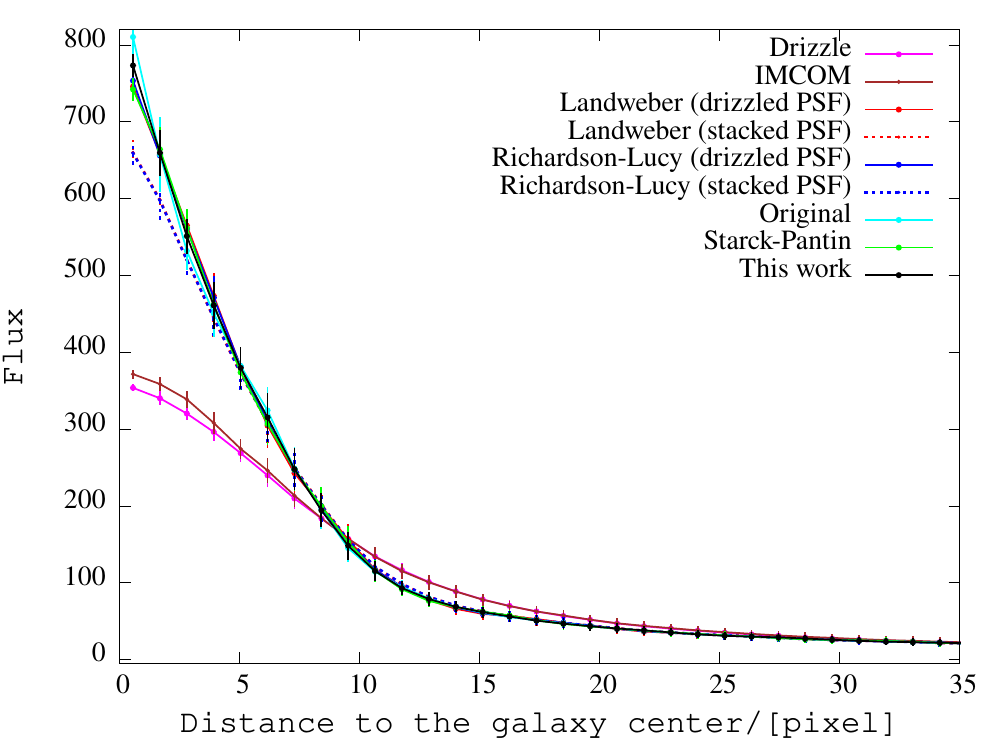} 
\caption{The ${\bf Gal-I}$ galaxy profiles from different recoveries.}
\label{GalaxyProfile}
\end{figure}

\begin{figure*}
\centering
\includegraphics[width=6in]{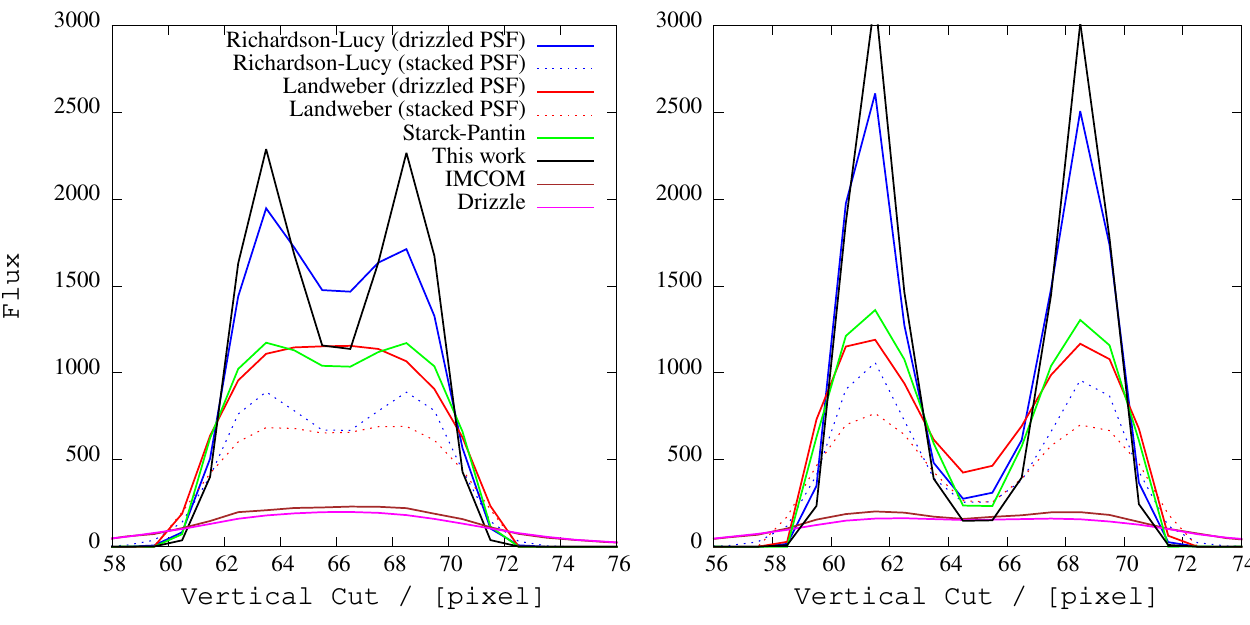} 
\caption{Super-resolution tests: a comparison of different recovery methods for two binaries (binary {\bf A} for left and binary {\bf B} for right panel). The vertical cut is a vertical section with respect to the pixel grid through the center of the binary members.}
\label{twins_peak}
\end{figure*}

\begin{figure*}
\centering
\includegraphics[width=7in,angle=0.0]{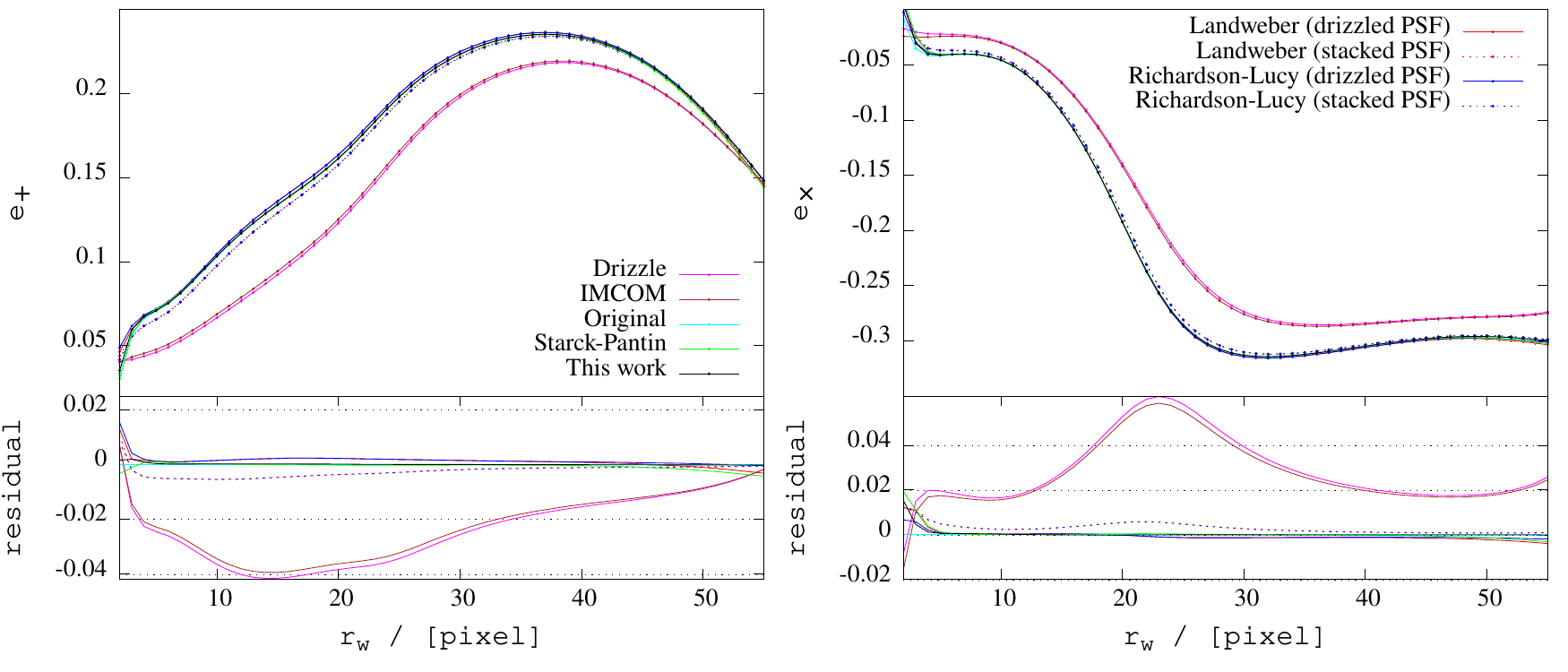} 
\caption{Shape parameters comparison between the original and recoveries: left for the $e_+$ and right for the $e_\times$. }
\label{shapefig}
\end{figure*}

\section{Simulation tests on the astronomical mock}\label{sec_ST}

In this simulation test, we use $Galsim$\cite[][]{Rowe2015} software\footnote{https://github.com/GalSim-developers/GalSim} to generate an HST Wide Field Camera 3 (WFC3) mock image with a size $401 \times 401$ pixels as the original (reference) image $f$ in which three samples of high-resolution HST galaxies with different profiles and intensities are introduced: ${\bf Gal-I, II\ \&\ III}$, as shown on the top left panel of Figure \ref{mockfig}. And we put an ideal star (one pixel\footnote{Here we do not convolve the ideal star with an HST PSF, because we want to obtain a {\it Drizzled} PSF.}, marked by a cross) on the top left to monitor the PSF. Three pairs of binary star systems, (i.e. binaries {\bf A}, {\bf B}, and {\bf C}) are put on the bottom right to check the super-resolution ability that different methods can achieve. 

If the PSF does not vary rapidly across the instrumental field of view, then images of different stars within a small region of a given exposure will approximate multiple frames of the same PSF, but with a variety of centroid offsets. Based on this assumption, 110 fine PSF kernels $\{b_1, b_2... b_{110}\}$\footnote{in $59\times 59$ pixels, must be in the same physical units as the original image $f$} are measured from the Very Large Telescope (VLT)\footnote{$r$-band VOICE-CDFS-1 exposures from the OmegaCAM of VLT Survey Telescope (VST) of European Southern Observatory.} sky field (centered on $R.A.=53^\circ.1511, DEC.=-27^\circ.7175$) by a Principal Component Analysis (PCA)-based method, which is developed by our collaborators \citet[][]{Nie2021a}. Stacking one by one pixel and normalizing the 110 PSF kernels $b_k$ forms a {\it stacked} PSF, as shown on the top left panel of Figure \ref{PSFs}.

Using the HST mock image $f$ and the measured PSFs $b_k$, we start mimicking the VLT exposures. After the PSF convolution, we bin the relatively shifted HST/WFC3 (fine) pixels by a down-sampling factor $\beta=0.21/0.04=5.25$\footnote{https://www.eso.org/sci/facilities/paranal/instruments/omegacam/inst.html \& https://esahubble.org/about/general/instruments/wfc3} to the VST/OmegaCAM (coarse) pixels. By adding noise measured from the same VLT sky field $\sigma_{\rm VLT}$\footnote{Note that $\sigma_{\rm VLT}$ is different for each frame.} to the coarse pixels, eventually, 110 VLT mock exposures $\{g_1, g_2... g_{110}\}$ ($115 \times 115$ pixels) are generated. One of them is shown on the top middle panel of Figure \ref{mockfig}. When the VLT multiple exposures are {\it Drizzled}, the ideal star forms a {\it Drizzled} PSF, as shown on the top left corner of {\it Drizzle} image (top right panel) in Figure \ref{mockfig}.

Namely, we generate the VLT mock frames from the high-resolution HST mock image. So that the HST/WFC3 mock image actually plays not only the role of the original image but also the role of the reference image in the simulation. Naturally, we can use a full-reference image quality assessment to determine the pseudo regularization parameter, i.e. the optimal number of iterations $O_{\rm iter}$ which has the highest PSNR among all the iterations (hereafter, all \texttt{iMECs} recoveries shown in visual are at the optimal iteration $O_{\rm iter}$). The iterative algorithms rely partly on circular convolutions computed through the Fast Fourier Transform, which actually results in the appearance of Fourier-related artifacts such as ghosts or ringings nearby the edge of the recovered image\citep[][]{Sage2017}, so that the PSNRs are calculated by using the pixels in the rectangles of ${\bf Gal-I,\ II}$ or ${\bf III}$ in Figure \ref{mockfig}.  

The $O_{\rm iter}$ actually acts as a tradeoff between image deblurring and noise amplification. Since we focus on the speed of convergence of different methods at the same step size parameter, in this article all $\gamma$s are set to be one in the following tests. Before running the simulation to determine the optimal number of iterations $O_{\rm iter}$, we list the image preprocessing recipe as follows.

\begin{enumerate}
\item The background correction: by definition, the recovery methods with a ratio correction require the flux with positive values on all involved pixels. Unfortunately, due to the background subtraction, there are a certain amount of pixels with negative flux in the frames, which has harmful effects on the recovery methods with a ratio correction, e.g. leading to high-frequency oscillations or even divergent results. Therefore, a temporary background correction is performed on the recoveries with a ratio correction: a positive constant background is added to the frames during the recovery and eventually removed from the recovered images. While this kind of temporary background correction is not necessary for the recovery methods with a difference correction. 

\item A projection on $\mathbb{R}^+$ (a non-strict positivity constraint) is put on all the recoveries to reduce the ringing effect:
\begin{equation}\label{cstr}
\Pp_{\mathbb{R}^+}\{ f^{(i+1)} \}=
\begin{cases}
f^{(i+1)}&f^{(i+1)} \ge 0\\
f^{(i)}&f^{(i+1)} < 0
\end{cases}
\end{equation}
where $\Pp_{\mathbb{R}^+}\{ f^{(i+1)} \}$ is a component-wise projection of $f^{(i+1)}$ onto the set $\mathbb{R}^+$ (not strictly). Compared with the constraint $\Pp_{\mathbb{R}^+}\{ f^{(i+1)} \}={\rm max}(f^{(i+1)},0)$ which is previously adopted in {\it SP02} and \citet[][]{Sage2017}, the equation \ref{cstr} can maintain the flux conservation and improve the fidelity of recoveries.

\end{enumerate}
This recipe is mainly aimed at background processing, which extends the range of the validity of the recoveries with a ratio correction, reduces the ringing effect, and suppresses the noise amplification.

\subsection{Up-sampling tests on the PSF reconstruction}\label{sec_DP}
We explore the fundamental properties of the up-sampling method using the VLT mock samples without PSF deconvolution. This allows us to perform an idealized test of the various up-sampling or anti-aliasing methods without complicating factors that may introduce their own sources of error. The equation \ref{model} shows that if we observe a point source by a telescope with CCD or CMOS acquisition system, the underlying PSF will be pixelated by the sampling, e.g. $\Ss^h\{\Bb\{\Ff\}\}(X_{m}, Y_{n})$. In this subsection, the underlying PSF can be regarded as a profile of an extended source. Sometimes, the spatial signal detection is aliased, i.e. the PSF is under-sampled. It is, therefore, necessary to construct an algorithm that provides accurate PSF kernels for the precision optimal photometry and morphology measurement. The PSF reconstruction becomes a project of up-sampling or anti-aliasing. Several methods are developed to reconstruct the underlying "unpixelized" optical PSF to super-resolution, e.g. as mentioned before, {\it Drizzle}\citep[][]{FH2002}, {\it Super-Drizzle}\citep[][]{Takeda2006},  \texttt{IMCOM}\citep[][]{Rowe2011}, {\it iDrizzle}\citep[][]{Fruchter2011}, {\it SPRITE}\citep[][]{Mboula2015}, {\it fiDrizzle}\citep[][]{WL2017}, and {\it IBP}\citep[][]{Irani1991,Symons2021}. Here we mainly compare the newly proposed approach {\it fiDrizzleRC} to the previous works {\it Drizzle}, {\it iDrizzle}, and \texttt{IMCOM}. The {\it iDrizzle} is often used to reduce the aliasing effect by coadding a series of under-sampled multi-exposures. And the \texttt{IMCOM} provides an oversampled output image from multiple under-sampled input images, assuming that the PSF is fully specified. 

We use {\it Drizzle}, {\it iDrizzle}, \texttt{IMCOM} and {\it fiDrizzleRC} to coadds 110 VLT mock frames $g_k$ in total. In the bottom array of Figure \ref{mockfig}, three recoveries from different methods are shown as \texttt{IMCOM}, {\it iDrizzle} and {\it fiDrizzleRC}. Visually, there is little difference between them. Then a zoomed-in version that focuses on the PSF is shown in Figure \ref{PSFs}. We find that comparing the {\it iDrizzle} and {\it fiDrizzleRC}, the \texttt{IMCOM} recovered PSF is slightly smooth (also see Figure \ref{PSFprofile}). Quantitatively, we plot the PSFs profiles and their differences (residuals) from the stacked PSF in Figure \ref{PSFprofile}. Basically, three up-sampling methods \texttt{IMCOM}, {\it iDrizzle} and {\it fiDrizzleRC} obtain sharp PSFs and approach the underlying one, i.e. the stacked PSF. Instead of the {\it Drizzled} PSF, it will be more accurate to construct a PSF model by using the up-sampled PSF.

\subsection{Implementation of \texttt{iMECs} on the VLT mocks}\label{sec_DPD}

Deblending in multi-exposures coaddition is a complex procedure that depends on accurate position calibration, frame weight, bad pixel mask, etc. In this work, we are only interested in achieving super-resolution by up-sampling and PSF deconvolution. Therefore, we assume that the position calibration, frame weight, etc. are well done. And the difference only results from the different choices of recovery methods. In order to provide a concrete test case on the up-sampling and PSF deconvolution, we implement six algorithms on the VLT mock sample as described above. Six recoveries from the \texttt{iMECs} are shown in Figure \ref{recoveryfig}, the top three with difference-corrections: {\it Landweber} and {\it Starck-Pantin} methods, while the bottoms with ratio-correction terms: {\it Richardson-Lucy} and {\it UPDC-RC} (i.e. this work). All of them are iterated from the same initial condition, the {\it Drizzled} image $f^{(0)}$, to the optimal number of iteration $O_{\rm iter}$ which is determined by the PSNR calculated from the pixels in the white rectangle (${\bf Gal-I}$) of Figure \ref{mockfig}. The {\it Landweber} (or {\it Richardson-Lucy}) with {\it drizzled PSF} represents one uses a {\it drizzled PSF} to deconvolve the initial image $f^{(0)}$.

The left panel of figure \ref{PSNRA} shows how the PSNR varies according to the iterations, from $\sim 39$ to $\sim 55$. Once the PSNR obtains its peak, the iteration stops. Basically, recoveries with ratio-correction terms converge more rapidly than that with difference-corrections. Among them, the {\it UPDC-RC} method achieves the highest PSNR and costs the least iterations in this test.

Although the non-strictly positive constraint is introduced to reduce the ringing, in Figure \ref{recoveryfig}, visually, the ringing effect can easily arise in the methods with difference-correction terms, e.g. {\it Landweber} with {\it (drizzled PSF)} and {\it Starck-Pantin} (see also the right panel in Figure \ref{starprofile}). Algorithms with ratio-corrections, such as {\it Richardson-Lucy} with {\it (drizzled PSF)} and {\it UPDC-RC}, can recover more accurately the point source (see also the left panel in Figure \ref{starprofile}) and extended source (Figure \ref{GalaxyProfile})  than the others. The deconvolutions with {\it stacked PSF} can not resolve the sources as well as others (see also Figure \ref{starprofile} and \ref{GalaxyProfile}). Compared with the Drizzled version on the top right panel in Figure \ref{mockfig}, all six recoveries illustrate the noise amplification. Fortunately, the signal extraction and noise amplification can be well-balanced by monitoring the PSNR, i.e. the optimal number of iterations $O_{\rm iter}$. Therefore, at least, at the $O_{\rm iter}$ iteration, the noise introduced by the recovery does not severely affect the flux and shape measurements on the stars and galaxies.

In order for the recoveries to check the effect quantitatively, we plot the profiles of the ideal star (the top left source on each panel) and the central galaxy in Figure \ref{starprofile} and \ref{GalaxyProfile} respectively. Obviously, {\it Richardson-Lucy} with {\it drizzled PSF} and {\it UPDC-RC} get sharper profiles and fewer ringings (see ringings around point sources on the top left and right panels) than the others. For extended sources, generally, the {\it UPDC-RC}, {\it Starck-Pantin} and deconvolutions with {\it drizzled PSF} can well reconstruct the central galaxy's profile. Since the \texttt{IMCOM} only deals with the up-sampling, it performs a little better than the {\it Drizzle}. While due to lack of pixelation information in the {\it stacked PSF}, the deconvolution with {\it stacked PSF} recovers a little worse than that with {\it drizzled PSF}.

The ability of super-resolution for different methods is shown as a vertical section in Figure \ref{twins_peak}. Stars in binary {\bf A} (or {\bf B}) is spaced by 5 (or 7) original (reference) pixels, corresponding to $5/\beta$ (or $7/\beta$) mock pixels. Therefore, the VLT mock sample is severely under-sampled. Figure \ref{twins_peak} illustrates that {\it Drizzle} and \texttt{IMCOM} fail to resolve binaries {\bf A} and {\bf B}, while all the other six recoveries can resolve binary {\bf B}. Among all the methods, only {\it UPDC-RC} can meet the Rayleigh criterion and resolve the two-point sources from binary {\bf A}.

To better understand to what extent the recovery methods can restore the galaxy morphology, we implement a shape test on the central spiral galaxy, which is often used in gravitational lensing measurement. For instance, in weak gravitational lensing studies, people are more concerned about the shear measurement, according to measuring the geometric distortion of the background galaxies over the randomly aligned ones, we can constrain the properties of the foreground lensing objects. Thus, the background galaxy shape i.e. the ellipticity is an important signal to calculate the gravitational lensing effect.

Following \citet[][]{HS2003}, the ellipticity of an object is defined as $e_+ = (M_{xx}-M_{yy})/(M_{xx}+M_{yy})$ and $e_\times = 2M_{xy}/(M_{xx}+M_{yy})$ where $M_{ij}$ represents the moments. the spin-2 tensor $\textbf{e} = (e_+,e_\times)$ is the so-called ellipticity tensor. In order to compare the methods quantitatively, we measure the geometric distortion of the galaxy for all recoveries. To avoid the problem of divergence, a circular Gaussian weighting function with a weight radius of $r_w$ is convolved into the galaxy, ${\bf Gal-I}$. The variation of shape parameters and their residuals (from the original) with respect to the weight radius $r_w$ is plotted in Figure \ref{shapefig}. 

According to Figure \ref{starprofile}, \ref{GalaxyProfile} and \ref{shapefig}, the advantages of the {\it UPDC-RC} are obvious in recovering the shapes and profiles of the star and galaxies at almost all radii. The largest deviation from the original occurs in the {\it Drizzle} and \texttt{IMCOM} recovered images. As for the traditional {\it stack-and-deconvolve} methods, compared to methods with {\it stacked PSF} deconvolution, those deconvolved by {\it drizzled PSF} perform better in the restoration of the galaxy morphology and star profile.

As discussed in Section \ref{sec_MEC}, the {\it drizzled PSF} evolving the pixelation blur is not homogenous on the image plane and should not be deconvolved as an ordinary PSF. However, if the number of exposures is very large, e.g. 110 frames, according to the appendix \ref{sec_APA}, the deconvolution with {\it drizzled PSF} converges towards the {\it UPDC-RC} and even the original image. Although the {\it Starck-Pantin} is theoretically reasonable in up-sampling and PSF deconvolution, it still suffers from the ringing effect to some extent and lacks the ability to obtain a higher PSNR and a sharper point source profile.

\subsection{Comparisons of simulations with different background noise}\label{sec_DPDC}
Here we regard the simulation in the above subsection as ${\bf Simu-1 }$. In order to test the reliability of different methods extensively, we run two other simulations with 4 and 8 times of $VLT$ background noise $\sigma_{\rm VLT}$ respectively, i.e. ${\bf Simu-2 }$ and ${\bf Simu-3 }$. Three simulations parameters setup is listed in table \ref{tab:params2}. They are obviously the same except for the background (Gaussian) noise amplitude. Therefore, we run three simulations: ${\bf Simu-1, 2\ \&\ 3.}$ for three galaxy stamps: ${\bf Gal-I, II\ \&\ III}$ to explore how the noise levels affect the target diagnostic quantities for different recoveries.
 \begin{table}
    \caption{Parameters used to generate ${\bf Simu-1, 2\ \&\ 3.}$}\label{tab:params2}
        \centering
        \begin{tabular}{c||ccc|}
        \diagbox{\bf{Simu No.}}{\bf{Params}} & {$\beta$}  & {\it Noise / [$\sigma_{\rm VLT}$]}  & Number of frames \\
        \hline
        1  & 2.0 & 1.0 & 110    \\
        2 & 2.0 & 4.0 & 110   \\
        3 & 2.0 & 8.0 & 110  
        \end{tabular}
    \end{table}

Besides the well-known diagnostic parameter PSNR which is used to evaluate the gray value similarity, we also analyze other two objective image quality metrics, the structural similarity index measure (SSIM), as well as the average deviation of all source fluxes ${\langle \delta F \rangle}$. Both of them are the full reference metric that requires two images from the same image capture $-$ a reference image $p$ and a recovered image $q$. 

 \begin{table*}
    \caption{The diagnostic parameters for different recovery methods and different sources. Boldface numbers indicate the best values in the recovery comparison. All the data are extracted from the optimal iterations.}\label{tab:params3}
        \centering
        \begin{tabular}{l||ccc|ccc|ccc|ccc|}
       &\multicolumn{3}{c|}{\it PSNR / {\rm [ dB ]}} &\multicolumn{3}{c|}{\it SSIM} &\multicolumn{3}{c|}{\it {$\langle \delta F \rangle$} / {\rm [ \% ] }} &\multicolumn{3}{c|}{$\it O_{\rm iter}$} \\
         \diagbox{\bf{Methods}}{\bf{Simulation No.}} & 1 & 2 & 3 & 1 & 2 & 3 & 1 & 2 & 3 & 1 & 2 & 3 \\
        \hline
        Gal-I Landweber (drizzled PSF) & 50.66 & 48.85 & 47.44 & 0.9921 & 0.9880 & 0.9836 & 0.0300 & 0.0433 & 0.2181 & 2366 & 749 & 291 \\
        Gal-I Landweber (stacked PSF) & 48.78 & 47.71 & 46.85 & 0.9871 & 0.9835 & 0.9799 & 0.0164 & 0.0270 & 0.1626 & 1914 & 446 & 198 \\
        Gal-I Starck-Pantin & 51.86 & 49.51 & 48.10 & 0.9940 & 0.9896 & 0.9856 & 0.0198 & 0.0580 & 0.0902 & 1162 & 314 & 139\\
        Gal-I Richardson-Lucy (drizzled PSF) & 53.76 & 51.87 & 50.01 & 0.9962 & 0.9940 & 0.9909 & 0.0492 & 0.0485 & 0.1830 & 909 & 320 & 122\\
        Gal-I Richardson-Lucy (stacked PSF) & 49.23 & 48.58 & 47.95 & 0.9884 & 0.9866 & 0.9844 & {\bf 0.0104} & {\bf 0.0116} & 0.1102 & 685 & 173 & 89\\
        Gal-I UPDC-RC (this work) & {\bf 54.76} & {\bf 52.15} & {\bf 50.57} & {\bf 0.9969} & {\bf 0.9943} & {\bf 0.9918} & 0.0142 & 0.0232 & {\bf -0.0739} & {\bf 546} & {\bf 153} & {\bf 76}\\
        \hline
        Gal-II Landweber (drizzled PSF) & 47.50 & 46.01 & 44.60 & 0.9892 & 0.9847 & 0.9792 & 0.0086 & 0.0204 & 0.3324 & 3079 & 1386 & 580 \\
        Gal-II Landweber (stacked PSF) & 45.46 & 44.56 & 43.87 & 0.9814 & 0.9772 & 0.9733 & 0.0084 & -0.0055 & 0.3226 & 2821 & 785 & 370 \\
        Gal-II Starck-Pantin & 49.32 & 46.90 & 45.39 & 0.9929 & 0.9875 & 0.9822 & -0.0027 & 0.0191 & 0.4264 & 2012 & 582 & 265\\
        Gal-II Richardson-Lucy (drizzled PSF) & 51.03 & 49.66 & 46.96 & 0.9952 & 0.9934 & {\bf 0.9877} & 0.0223 & 0.0095 & 0.2630 & 1019 & 504 & 151\\
        Gal-II Richardson-Lucy (stacked PSF) & 45.82 & 45.34 & 44.89 & 0.9829 & 0.9809 & 0.9788 & 0.0041 & {\bf -0.0041} & 0.2563 & 940 & 242 & 135\\
        Gal-II UPDC-RC (this work) & {\bf 52.79} & {\bf 50.01} & {\bf 46.98} & {\bf 0.9968} & {\bf 0.9939} & 0.9874 & {\bf 0.0021} & 0.0107 & {\bf 0.2321} & {\bf 861} & {\bf 239} & {\bf 39}\\
        \hline
        Gal-III Landweber (drizzled PSF) & 33.61 & 32.34 & 31.02 & 0.9760 & 0.9674 & 0.9562 & -0.0245 & 0.1537 & 0.3579 & 185 & 30 & 11 \\
        Gal-III Landweber (stacked PSF) & 33.42 & 32.21 & 30.87 & 0.9742 & 0.9657 & 0.9533 & -0.0441 & 0.0636 & 0.2138 & 121 & 24 & 10 \\
        Gal-III Starck-Pantin & 33.82 & 32.47 & 31.00 & 0.9770 & 0.9682 & 0.9552 & -0.0437 & 0.0863 & 0.2296 & 95 & 20 & 9\\
        Gal-III Richardson-Lucy (drizzled PSF) & 33.85 & 32.69 & 31.37 & 0.9773 & 0.9699 & 0.9597 & -0.0254 & 0.1147 & 0.1516 & 146 & 24 & 10\\
        Gal-III Richardson-Lucy (stacked PSF) & 33.56 & 32.50 & 31.14 & 0.9751 & 0.9679 & 0.9562 & -0.0468 & {\bf 0.0266} & {\bf -0.0103} & 98 & 20 & 9\\
        Gal-III UPDC-RC (this work) & {\bf 34.04} & {\bf 32.79} & {\bf 31.39} & {\bf 0.9969} & {\bf 0.9705} & {\bf 0.9606} & {\bf -0.0229} & 0.0294 & -0.0342 & {\bf 82} & {\bf 17} & {\bf 8}\\

        \end{tabular}
    \end{table*}

The SSIM is a perceptual metric that quantifies image quality degradation caused by source distortion, variation of brightness, and contrast. It is mainly employed to reflect the structural similarity. Following \citet[][]{Hore2010}, the SSIM is defined as:
\begin{equation}\label{SSIM}
SSIM ( p , q ) = l ( p , q ) c ( p , q ) s ( p , q ) ,
\end{equation}
where
\begin{equation}\label{SSIMP}
\begin{dcases}
l( p , q ) =& \frac{2\mu_p\mu_q+C_1}{\mu_p^2+\mu_q^2+C_1}\\
c( p , q ) =& \frac{2\sigma_p\sigma_q+C_2}{\sigma_p^2+\sigma_q^2+C_2}\\
s( p , q ) =& \frac{2\sigma_{pq}+C_3}{\sigma_p\sigma_q+C_3}
\end{dcases}
\end{equation}
The first term in equation \ref{SSIMP} is the luminance comparison function which measures the closeness of the two images' mean luminance ($\mu_p$ and $\mu_q$). The second term is the contrast comparison function which measures the closeness of the contrast between the two images. The contrast is measured by the standard deviation $\sigma_p$ and $\sigma_q$. The third term is the structure comparison function which measures the correlation coefficient between the two images $u$ and $v$. And $\sigma_{pq}$ is the covariance between $u$ and $v$. The positive constants $C_1$, $C_2$ and $C_3$ are used to avoid a null denominator. Here we fix the constants to $C_3=C_2/2$ and $C_2=C_1=0.1$.

Following \citet[][]{Symons2021}, we define the average deviation of all source fluxes ${\langle \delta F \rangle}$ determined via optimal photometry\citep[][]{Naylor1998} from known values to be
\begin{equation}
\label{eq:fdev}
{\langle \delta F \rangle} = \frac{\sum\limits_{i}^{N} [(F_{\rm{P},\it{i}}-F_{i})/F_{i}]}{N},
\end{equation}
where $\langle \delta F \rangle$ is expressed in percent, $N$ is the number of sources in the image, $F$ is a source's known flux in the reference image, and $F_{\rm P}$ is that computed by optimal photometry in the recovered image. $\langle \delta F \rangle$ is an important index to assess the reliability of a recovery method, especially for astronomical photometry.

The diagnostic parameters for the optimal iterations of three simulations are illustrated in table \ref{tab:params3}. The $\it O_{\rm iter}$s for different galaxy stamps are determined by the PSNRs in the rectangles of ${\bf Gal-I,\ II}$ or ${\bf III}$ in Figure \ref{mockfig}. Basically, the trend of SSIM follows the PSNR,  their values decrease with the increase of the background noise. And so does the $\it O_{\rm iter}$. The higher amplitude of background noise, the earlier the noise amplification dominates in the recovery iterations. The $\langle \delta F \rangle$ increases when the noise of the background arises. {\it UPDC-RC} and {\it Richardson-Lucy} with {\it stacked PSF} have better performance in reducing the average deviation of all source fluxes $\langle \delta F \rangle$. Galaxies ${\bf Gal-I}$ and ${\bf II}$ have similar intensities and profiles, then have similar diagnostic parameters. While stamp ${\bf Gal-III}$ is very different from ${\bf Gal-I}$ and ${\bf II}$, which eventually results in a big gap of diagnostic parameters to those of ${\bf Gal-I}$ or ${\bf II}$.

In order to quantify the deviation of the recovered extended sources (e.g. galaxies) from the reference ones, we estimate the source distortion ratio (SDR) of the reconstruction and the $\chi^2$ for galaxy ellipticity ($e_+, e_\times$) and profile (flux). Following \citet[][]{Joseph2021}, for each recovered image, we compute the source distortion ratio of the recovery defined as:
    \begin{equation}\label{eq:SDR}
        SDR(q) = 10\log_{10}\Big{(}\frac{||p||}{||q - p||} \Big{)},
    \end{equation}
    where $q$ is the recovered image, $p$ is the reference image, and $||.||$ is the Euclidean 2-norm. The higher the SDR, the more accurate recovery is. The $\chi^2$ for shape parameters is defined as:
    \begin{equation}\label{eq:CHI2}
        \chi^2(\tilde{t}) = \sum\limits_{i}^{N}\frac{(\tilde{t_i}-t_i)^2}{t_i},
    \end{equation}
  where variate $t$ can be replaced by the galaxy ellipticity ($e_+, e_\times$) or flux. $\tilde{t_i}$ is calculated from the recovered image, such as the points in figure \ref{shapefig} or \ref{GalaxyProfile}, $t_i$ is calculated from the original (reference) image.
  
\begin{figure*}
\centering
\includegraphics[width=7in,angle=0.0]{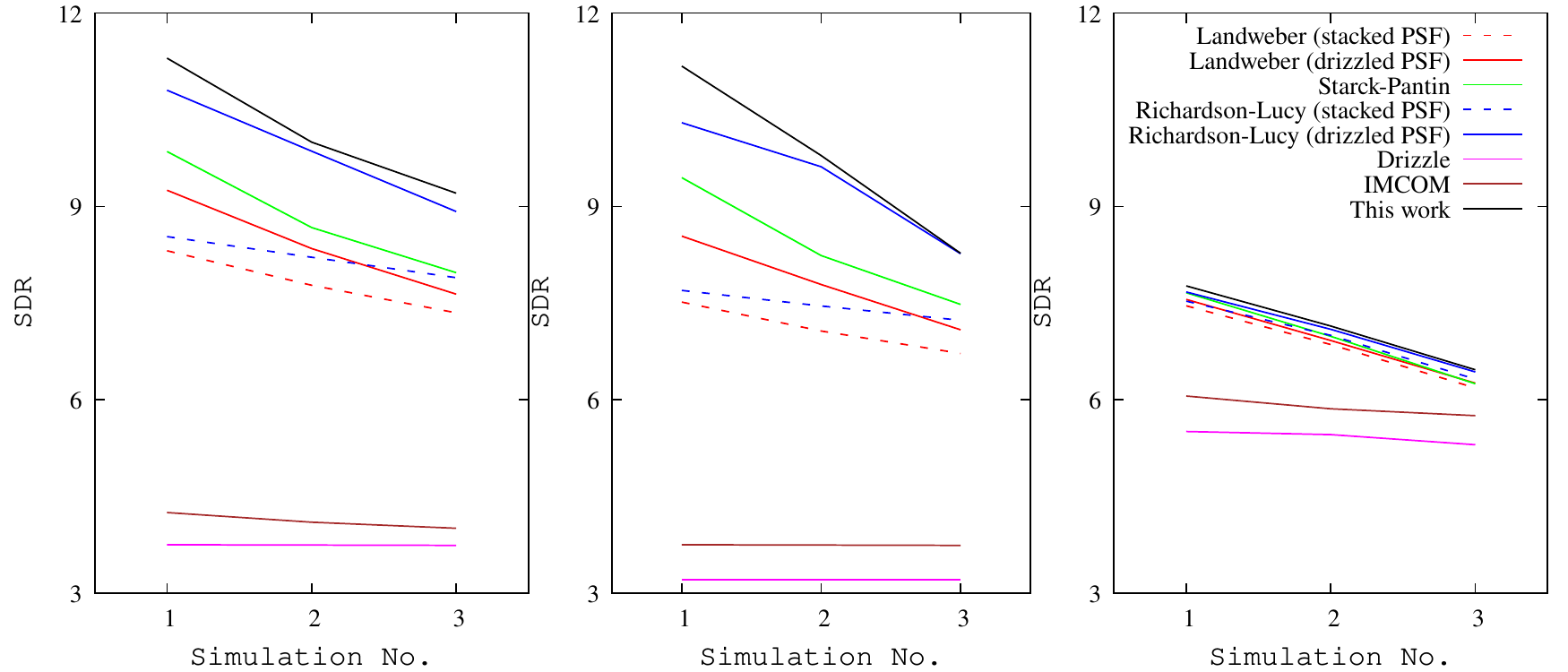} 
\caption{The SDRs for three galaxy stamps (left panel: ${\bf Gal-I}$, middle: ${\bf Gal-II}$, right: ${\bf Gal-III}$ ) recovered by different methods in the three simulations ${\bf Simu-1, 2\ \&\ 3.}$}
\label{SDR}
\end{figure*}

\begin{figure*}
\centering
\includegraphics[width=7in,angle=0.0]{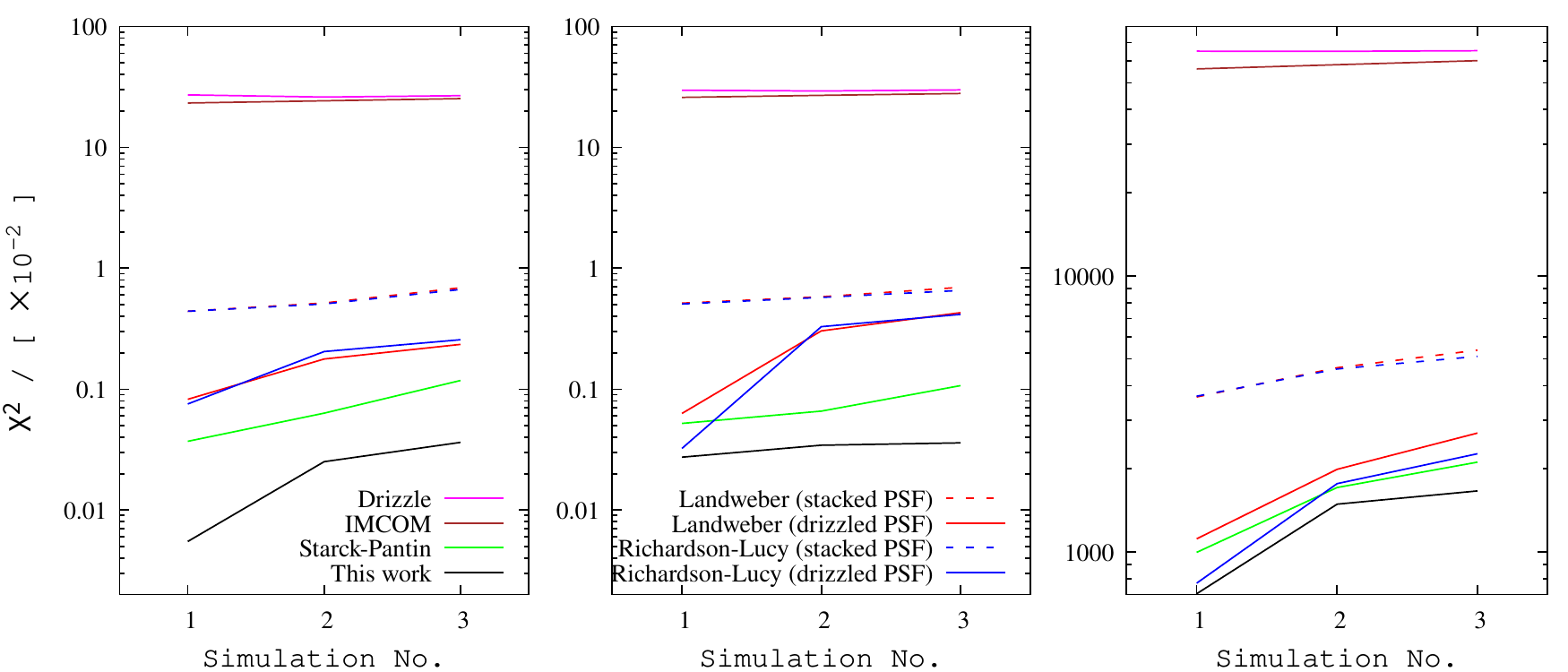} 
\caption{The $\chi^2$ for galaxy ${\bf Gal-I}$: left panel for the ellipticity parameter $e_+$, middle for $e_\times$ and right for the galaxy profile (flux). }
\label{CHI2}
\end{figure*}

In figure \ref{SDR}, we show the SDRs of the recovered images as a function of the sequence number of simulations for three stamps ${\bf Gal-I, II\ \&\ III}$. Generally, the SDR reduces as the amplitude of background noise increases. {\it UPDC-RC} and {\it Richardson-Lucy} with {\it drizzled PSF} have better performance than the other methods in all the galaxies. The $\chi^2$s of three parameters ($e_+, e_\times$ and flux, only for galaxy ${\bf Gal-I}$) in figure \ref{CHI2} show that {\it UPDC-RC} provides the best recoveries for all the three simulations. Figure \ref{SDR} and \ref{CHI2} illustrate that the iMEC methods are more sensitive to background noise than the linear methods e.g. {\it Drizzle} and \texttt{IMCOM}. However, {\it Drizzle} and \texttt{IMCOM} loose the most fidelity among all the recovery algorithms and galaxy stamps.
  
\section{Discussion and Conclusion}\label{sec_DAC}
In this article, we introduce an up-sampling method, {\it fiDrizzleRC}, and an alternative up-sampling and PSF deconvolution algorithm, {\it UPDC-RC}. Both {\it fiDrizzleRC} and {\it UPDC-RC} have ratio-correction terms and perform well in simulation tests. Especially, in the VLT mock tests, comparing to the previous works, e.g. the {\it stack-and-deconvolve} methods: {\it Landweber} and {\it Richardson-Lucy}, the optimal linear method: \texttt{IMCOM} and the up-sampling and the PSF deconvolution algorithm: {\it Starck-Pantin}, the new method {\it UPDC-RC} maintains competitive advantage in point and/or extended sources reconstruction, such as super-resolution, PSNR, SSIM, ${\langle \delta F \rangle}$, SDR and $\chi^2$. {\it Starck-Pantin} and {\it UPDC-RC} have the ability to handle up-sampling and PSF deconvolution simultaneously and improve the under-sampled multiple exposures to a super-resolution grid. There are still some details to discuss as follows.
\begin{enumerate}
\item Theoretically, it is not recommended to directly deconvolve PSF on the drizzled image by using the {\it drizzled PSF} (like the {\it stack-and-deconvolve} with {\it drizzled PSF}). Because the {\it drizzled PSF} includes the inhomogeneous pixelation blur which could not be treated as the ordinary PSF unless the number of coadded exposures is sufficiently large as mentioned in Section \ref{sec_MEC} and Appendix~\ref{sec_APA}. 

\item However, in practice, {\it stack-and-deconvolve} is an alternative option when the multi-exposures are adequate. The {\it stack-and-deconvolve} methods need only once to measure the PSF on the drizzled image, e.g. the {\it drizzled PSF}, while the {\it UPDC} needs to measure the PSF for each exposure. So {\it stack-and-deconvolve} saves more computation. Nevertheless, one should carefully use the {\it stack-and-deconvolve} algorithm because it maybe brings a significant ringing effect into the output.

\item Since the {\it stacked PSF} does not contain any information about the pixelation blur, the {\it stack-and-deconvolve} with a {\it stacked PSF} should not be used in the PSF deconvolution with a down-sampling factor $\beta >1$ or shifts $\Delta_k \neq 0$ in principle. Sometimes, the {\it stack-and-deconvolve} with {\it stacked PSF} can be regarded as a lower limit reference for the PSF deconvolution of the multi-exposure coaddition. Moreover, it performs well in the test of the average deviation of all source fluxes $\langle \delta F \rangle$ (see table \ref{tab:params3}). The {\it stack-and-deconvolve} with {\it stacked PSF} has similarity with the method proposed by \citet[][]{Hunt1994}, which introduces the interpolation to the PSF deconvolution (the PSF must be sampled on the same grid as the object).

\item Like the ringing effect, the noise amplification effect also depends on the number of iterations. In this work, we adopt the PSNR as an indicator to search the pseudo regularization parameter i.e. $O_{\rm iter}$. And by this way, the signal reconstruction and the noise amplification are well balanced. One can try other quality assessment parameters, like SSIM or SDR, according to his/her points of focus.

\item It is worth noting that owing to the PSF convolution and resampling executed in each exposure, {\it Starck-Pantin} and {\it UPDC-RC} methods consume much more computation than the {\it stack-and-deconvolve}. As for the example in Figure \ref{recoveryfig}, the {\it UPDC-RC} methods spend approximately 9 times longer than the {\it stack-and-deconvolve}, e.g. the {\it Landweber}, to finish the iteration.

\item In this article, we coadd multiple exposures from the same instrument for simplicity, then all down-sampling factors $\beta_k$ are the same. In fact, the \texttt{iMECs} is able to coadd the exposures from different telescopes like \citet[][]{Joseph2021} have done in their work.

\end{enumerate}

After a series of formula derivations and a set of numerical simulation tests, we conclude this work as follows.
\begin{enumerate}

\item We provide an alternative algorithm {\it fiDrizzleRC} (equation  \ref{GWLM}) to up-sample the under-sampled multi-exposures. As shown in the simulation test, {\it fiDrizzleRC} performs as well as the previous works \texttt{IMCOM} and {\it iDrizzle}. 

\item Besides the previous method {\it Starck-Pantin} (equation \ref{WLMPHDC}), we also develop {\it UPDC-RC} (equation  \ref{WLMPHRC}) as a supplement to the \texttt{iMECs}, which has a ratio-correction term. As shown in Section \ref{sec_DPD}, the {\it UPDC-RC} has better performance than the  {\it Starck-Pantin} and all of the traditional {\it stack-and-deconvolve} methods in the simulation tests. Especially, {\it UPDC-RC} is the only method that can resolve the binary {\bf A}. Maybe, given more iterations to the {\it stack-and-deconvolve} methods, they can also meet the Rayleigh criterion. However, the more iterations beyond the $O_{\rm iter}$, the stronger noise will be introduced.

\item Different from the traditional {\it stack-and-deconvolve} with {\it drizzled PSF} which actually employs a pseudo-up-sampling mechanism (e.g. the pixelation blur problem), the {\it Starck-Pantin} and {\it UPDC-RC} demonstrate a real sense of up-sampling and PSF deconvolution process in both theory (equation \ref{WLMPHDC}) and practice (Section \ref{sec_DPD} and  \ref{sec_DPDC}).

\end{enumerate}

Several iMEC methods, e.g., {\it iDrizzle}, {\it fiDrizzle}, {\it fiDrizzleRC}, {\it Landweber}, {\it Richardson-Lucy}, {\it Starck-Pantin} and {\it UPDC-RC}, are now packaged as a software module, \texttt{iMECs}, which is designed to coadd thousands of multi-exposures from the Multi-channel Imager (MCI) of Chinese Survey Space Telescope (CSST). 

\section*{Acknowledgements}
We thank the anonymous referee for helpful comments that greatly improved the presentation of this paper. We also thank Dezi Liu (South-Western Institute for Astronomy Research, Yunnan University, Kunming, China) for kindly providing the VLT frames, Lin Nie (Wuhan Institute of City, Wuhan, China) for measuring the high-resolution PSFs of VLT frames used in this article, and Huanyuan Shan (Shanghai Astronomical Observatory, Shanghai, China) for providing computation resources and program optimization suggestions. L.W. acknowledges the cosmology simulation database in the NBSDC. This work is also supported by the Foundation for Distinguished Young Scholars of Jiangsu Province (No. BK20140050), the National Natural Science Foundation of China (Nos. 11973070, 11333008, 11273061, 11825303, and 11673065), the China Manned Space Project with No. CMS-CSST-2021-A01, CMS-CSST-2021-A03, CMS-CSST-2021-B01, Key Research Program of Frontier Sciences, CAS, Grant No. ZDBS-LY-7013 and the Joint Funds of the National Natural Science Foundation of China (No. U1931210).

\begin{figure*}
\centering
\includegraphics[width=7in]{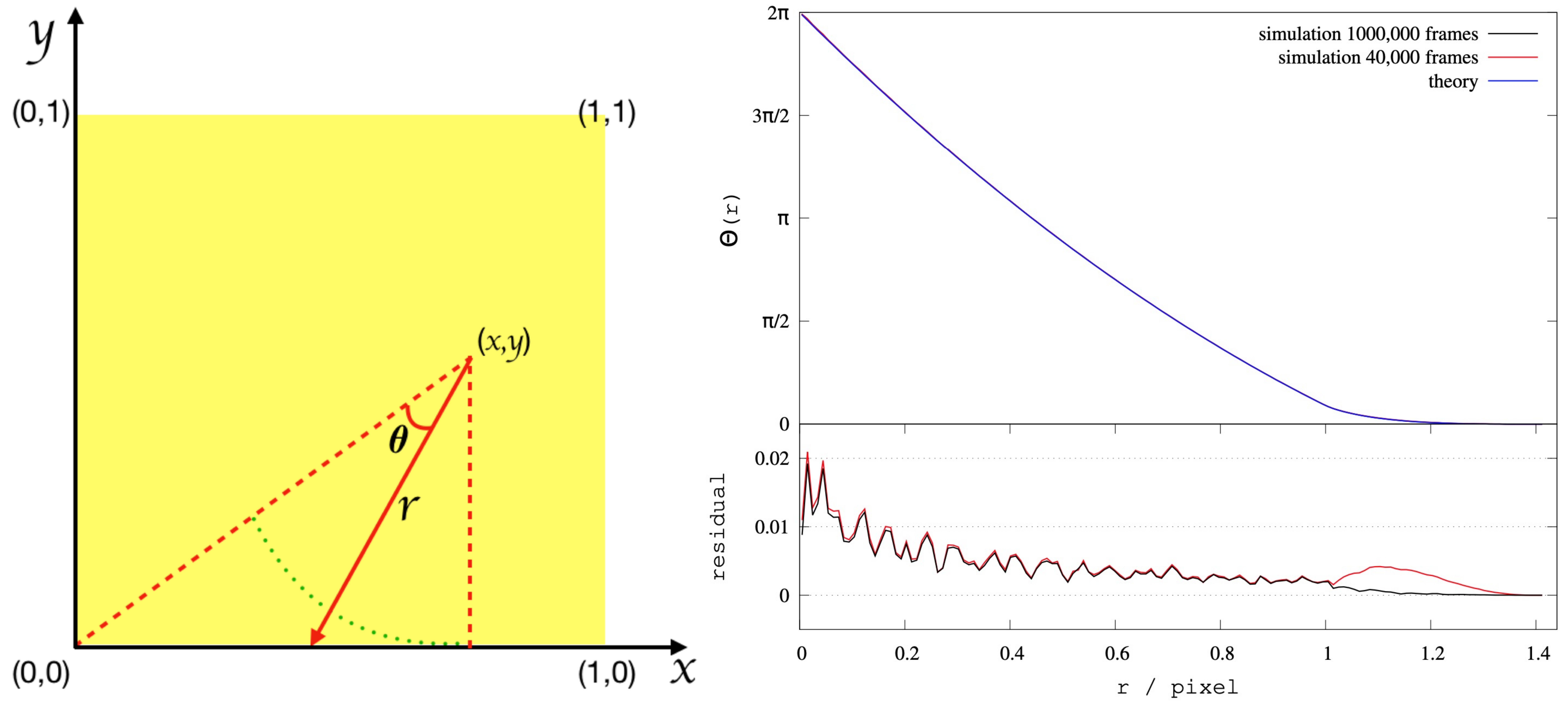} 
\caption{A sketch for calculating the thickness distribution of note papers (left panel) and a comparison between theory and the simulations (right panel).}
\label{APD}
\end{figure*}

\section*{Data Availability}
The data underlying this paper will be shared upon reasonable request to the corresponding author.



\bibliographystyle{mnras}
\bibliography{example} 




\appendix
\section{The note spike profile}
\label{sec_APA}
Without the PSF, a point source will hit only one pixel of the CCD. If coadding the multiple exposures of a point source, one will get a bright spot, i.e. pixelation blur. When both the number of the multiple exposures and the down-sampling factors are sufficiently large i.e. $\beta \gg 1$, there is a special form for the bright spot which is independent of position. Figuratively speaking, a fine target cell that encloses the point source is likened to a note spike, while the coarse pixels with flux on multiple exposures are likened to the note papers (assuming a square shape). The normalized thickness distribution of note papers has the same form as the profile of the bright spot -- we call it a note spike profile. In the sketch Figure \ref{APD} (left panel), any point $(x,y)$\footnote{$0\le x\le y\le1$} is the position of the spike, $r$ is the distance to the spike. The thickness contribution of a note paper can be written as
\begin{equation}
\theta(x, y, r)=
\begin{cases}
\displaystyle{\rm arctan}\frac{x}{y}&r < y\\
\displaystyle{\rm arctan}\frac{x}{y}-{\rm arccos}\displaystyle\frac{y}{r}&y \le r < \sqrt{x^2+y^2}\\
0&\sqrt{x^2+y^2} \le r
\end{cases}
\end{equation}
Obviously, due to the symmetry of the square note paper, there are 8 identical regions of integration in total. Therefore the final expression of the note spike profile is
$\Theta(r)=8\times\int_0^1\int_0^1 \theta(x, y, r){\rm d}x {\rm d}y$, i.e.


\begin{equation}\label{eq_NSP}
\Theta(r)=
\begin{cases}
2r^2-8r+2\pi&0\le r < 1\\
\displaystyle8{\rm arcsin}\frac{1}{r}-2r^2+8\sqrt{r^2-1}-4-2\pi &1 \le r < \sqrt{2}\\
0&else
\end{cases}
\end{equation}

In order to test the theory, we run two simulations with $\beta$=200 (40,000 exposures) and  $\beta $=1,000 (1,000,000 frames). The profiles are illustrated on the right panel of Figure \ref{APD}. The maximum deviation between the simulations and the theory is less than $0.8\%$ when the radius $r\le 1$. In this sense, one can anti-alias the under-sampled multi-exposures by deconvolving the {\it Drizzled} image with a normalized note spike profile. So it will be a shortcut to constructing the underlying PSF. Furthermore, the {\it drizzled PSF} can be approximatively replaced by a convolution between a normalized note spike profile and a {\it stacked PSF}. 



\bsp	
\label{lastpage}
\end{document}